\documentclass[letterpaper]{article} 
\usepackage{aaai2027}  
\usepackage{times}  
\usepackage{helvet}  
\usepackage{courier}  
\usepackage[hyphens]{url}  
\usepackage{graphicx} 
\usepackage{natbib}  
\usepackage{caption} 
\usepackage{algorithm}
\usepackage[noend]{algpseudocode}

\usepackage{newfloat}
\usepackage{listings}
\DeclareCaptionStyle{ruled}{labelfont=normalfont,labelsep=colon,strut=off} 
\floatstyle{ruled}
\newfloat{listing}{tb}{lst}{}
\floatname{listing}{Listing}
\usepackage{booktabs}
\usepackage{array}
\usepackage{multirow}
\usepackage{subfigure}

\usepackage{amsmath}  
\usepackage{amsfonts}
\usepackage{enumitem}
\usepackage{tabularx}
\usepackage[english]{babel}
\usepackage{amsthm}
\usepackage{bm}

\newcommand{\eg}{\emph{e.g., }}

\newcommand{\tablestyle}[2]{\setlength{\tabcolsep}{#1}\renewcommand{\arraystretch}{#2}\centering\footnotesize}

\floatname{algorithm}{Algorithm}

\algrenewcommand{\algorithmicrequire}{\textbf{Input:}}
\algrenewcommand{\algorithmicensure}{\textbf{Output:}}

\usepackage{times}
\usepackage{latexsym}

\usepackage[T1]{fontenc}

\usepackage[utf8]{inputenc}

\usepackage{microtype}

\usepackage{inconsolata}

\usepackage{graphicx}

\usepackage{xcolor}
\usepackage{pifont}       
\usepackage{bbding}       
\usepackage{fontawesome}  
\usepackage[noabbrev,capitalise]{cleveref}
 
\newcommand{\modelname}{ACERec}
\newcommand{\fakeparagraph}[1]{\noindent\textbf{#1.}}

\nocopyright 

\title{Unleash the Potential of Long Semantic IDs for Generative Recommendation}
\author{
Ming Xia\textsuperscript{1}\thanks{Equal contribution.},
Guoxin Ma\textsuperscript{2}\footnotemark[1],
Zhiqin Zhou\textsuperscript{3}\footnotemark[1],
Dongmin Huang\textsuperscript{1}\thanks{Corresponding author.}
}

\affiliations{
\textsuperscript{1}Southern University of Science and Technology, 
\textsuperscript{2}Xi'an Jiaotong University,
\textsuperscript{3}Nanjing University\\
{
\{xiamingtx03, guoxinma9, zinzhou0110, dmhuang777\}@gmail.com
}
}

\usepackage{bibentry}

\begin{document}

\maketitle

\begin{abstract}
Semantic ID-based generative recommenders face a granularity-efficiency dilemma between efficient recommendation with short IDs and expressive item modeling with long IDs.
To break this dilemma, we propose \modelname{}, a framework that preserves the semantic richness of long IDs while keeping the recommendation process efficient.
Concretely, \modelname{} employs an Attentive Token Merger to compress long semantic IDs into compact yet faithful latent tokens. To better capture user intent from the compressed semantics, we further introduce a dedicated Intent Token, optimized by a dual-granularity objective that combines token-level generation with item-level intent-semantic alignment. 
Extensive experiments on nine real-world benchmarks show that \modelname{} consistently outperforms state-of-the-art methods, yielding average relative improvements of 12.92\% in NDCG@10 and 7.49\% in Recall@10 over the strongest baselines.
\end{abstract}


\section{Introduction}\label{sec:introduction}
Generative recommendation has recently emerged as a promising alternative to the conventional ``retrieve-then-rank'' pipeline~\cite{wang2023generative, wang2025generative, li2024large}. A central design choice is how items are represented. Item ID-based recommendation typically represents each item with a unique ID and an independently learned embedding~\cite{kang2018self, zhou2020s3}. Although simple and effective, such representations encode little semantic information about the item itself. As a result, items with similar attributes are still modeled as unrelated IDs, limiting cross-item knowledge sharing and leaving items with sparse interactions poorly represented. Generative recommendation alleviates this limitation by representing each item as a sequence of semantic tokens drawn from a shared vocabulary~\cite{rajput2023recommender, hou2023learning}. Since related items can share tokens that reflect common attributes, the model can transfer knowledge across items and predict future interactions in a more scalable manner.

\begin{figure}[t]
    \centering
    \includegraphics[width=\linewidth]{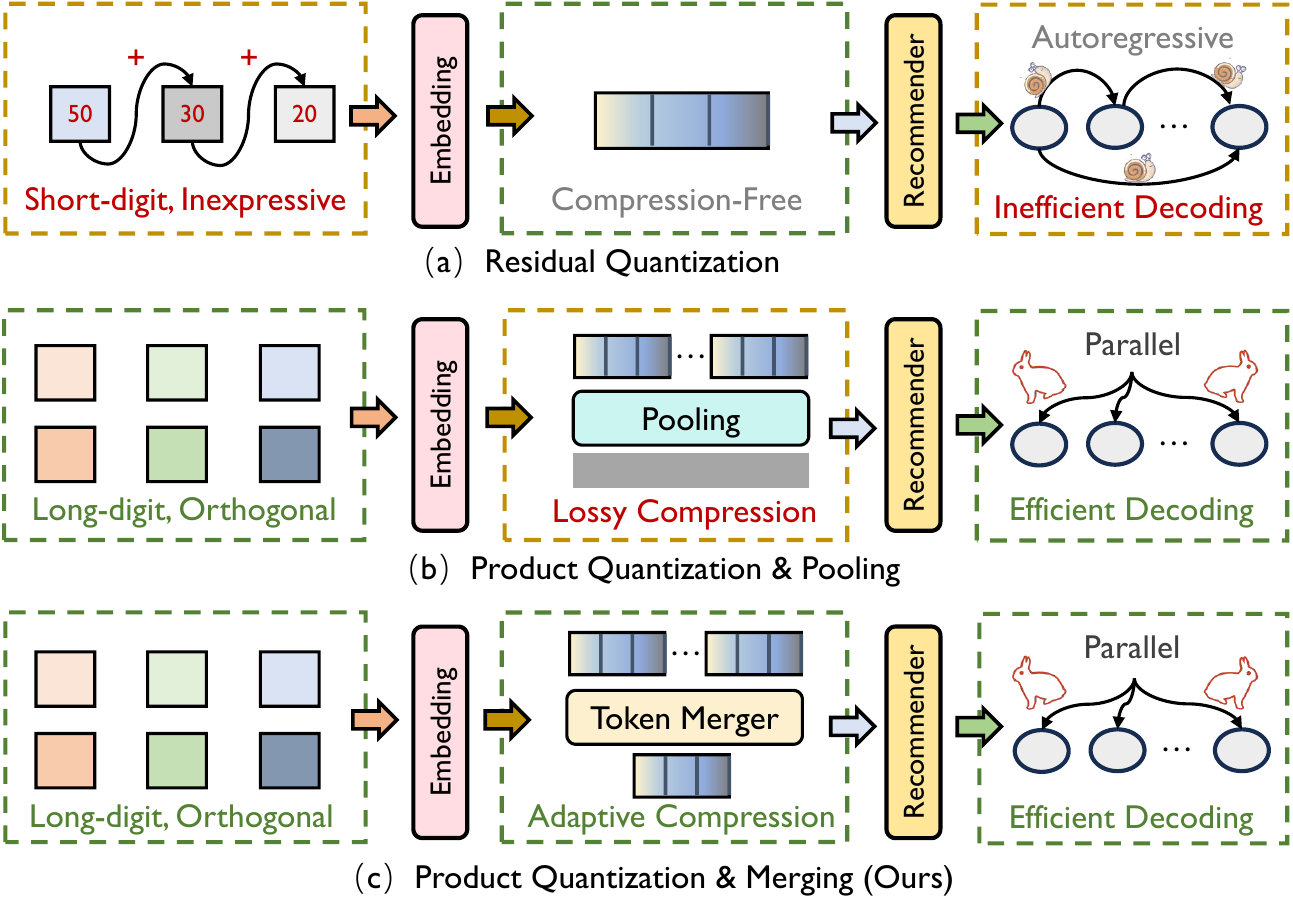}

    \caption{\textbf{Comparison of semantic-ID paradigms.}
(a) RQ: short hierarchical IDs with sequential decoding.
(b) OPQ + mean pooling: long parallel IDs compressed into a single vector.
(c) ACERec: long parallel IDs adaptively compressed into compact latent tokens.}
    \label{Fig: abstract}
\end{figure}

Existing generative recommenders mainly follow two tokenization paradigms.
As illustrated in Figure~\ref{Fig: abstract}(a), Residual Quantization (RQ)-based methods progressively quantize residual semantics into hierarchical IDs. However, the resulting codes are sequentially dependent, making long IDs costly to decode, while later codes mainly capture increasingly weak residual information and are difficult to fully exploit. In contrast, Optimized Product Quantization (OPQ)-based methods in Figure~\ref{Fig: abstract}(b) decompose item embeddings into orthogonal subspaces and quantize them independently~\cite{ge2013optimized,jegou2010product,hou2025generating}. The resulting long IDs capture complementary semantics from multiple subspaces and support parallel prediction. Nevertheless, directly modeling these tokens substantially lengthens user sequences and incurs prohibitive memory costs. Existing methods therefore rely on mean pooling~\cite{hou2023learning,hou2025generating}, which blurs fine-grained subspace semantics and weakens the value of long IDs. This raises a critical question: \textit{``Can we preserve the expressive potential of long semantic IDs while maintaining efficient input modeling and output prediction?''} We argue that the key is to decouple tokenization granularity from recommendation complexity, allowing the tokenizer to retain rich semantics while the recommender operates on compact yet faithful representations.

Based on this insight, we propose \modelname{} (\textbf{A}daptive \textbf{C}ompression for \textbf{E}fficient \textbf{Rec}ommendation), a framework that preserves the expressiveness of long semantic IDs while enabling efficient sequential modeling. Specifically, \modelname{} introduces an Attentive Token Merger (ATM) to adaptively distill long semantic sequences (e.g., $m=32$) into a small set of faithful latent tokens (e.g., $k=4$), allowing the recommender to exploit fine-grained item semantics without directly processing prohibitively long inputs. Building on these compact semantic representations, \modelname{} further introduces an Intent Token as a context-conditioned prediction anchor to capture the user's evolving preference from historical interactions. A dual-granularity objective jointly supervises fine-grained semantic ID prediction and Intent-Semantic Alignment, enabling accurate semantic ID generation while aligning the learned intent representation with the target item's holistic semantics. 
Our contributions are summarized as follows:
\begin{itemize}

    \item We propose \modelname{}, which decouples semantic tokenization granularity from sequential modeling, preserving the expressive power of attribute-rich long semantic IDs while maintaining tractable computation.

    \item We introduce a context-conditioned Intent Token to capture evolving user preferences, together with a dual-granularity objective that combines fine-grained semantic ID prediction with item-level intent--target alignment.

    \item Extensive experiments on nine real-world benchmarks demonstrate that \modelname{} consistently achieves state-of-the-art performance and exhibits strong knowledge transfer to sparse and cold-start items.

\end{itemize}
\section{Related Work}\label{sec:related_work}
\fakeparagraph{Generative Sequential Recommendation} Sequential recommendation has evolved from transition-based Markov Chains~\cite{rendle2010factorizing} and temporal RNNs/CNNs~\cite{hidasi2015session, li2017neural, tang2018personalized, yue2024linear} to modern self-attentive Transformers like SASRec~\cite{kang2018self} and BERT4Rec~\cite{sun2019bert4rec}. However, traditional models represent items as atomic IDs, leading to large catalog-dependent embedding tables and severe sparsity-induced information isolation~\cite{zhou2020s3}, even with auxiliary features~\cite{zhang2019feature}. To address this, generative recommendation~\cite{geng2022recommendation, rajput2023recommender} maps items into sequences of discrete semantic tokens within a shared vocabulary. Yet, it faces a fundamental trade-off between expressiveness and efficiency: RQ-based methods~\cite{rajput2023recommender, liu2025generative} restrict code lengths to bypass decoding bottlenecks but sacrifice resolution, whereas OPQ-based models~\cite{hou2023learning, hou2025generating} leverage orthogonal subspaces for attribute-rich long IDs (e.g., 32 tokens) but trigger quadratic attention overhead. Existing solutions control sequence length through rigid pooling, which uniformly aggregates complementary subspace semantics and weakens their utility for downstream recommendation. \modelname{} resolves this limitation by decoupling tokenization granularity from sequential modeling, preserving fine-grained semantics while maintaining efficient computation.


\fakeparagraph{Intent Modeling and Dual-Granularity Alignment}
To capture latent user preferences, intent learning has evolved from multi-interest networks~\cite{li2019multi, cen2020controllable} based on dynamic routing or attention to contrastive intent learning methods such as ICLRec~\cite{chen2022intent} and ELCRec~\cite{liu2024end}. Despite these advances, existing intent models typically rely on holistic item representations and therefore make limited use of the fine-grained semantics provided by semantic IDs. Meanwhile, semantic ID-based recommenders mainly focus on predicting individual semantic tokens, leaving their potential for modeling evolving user preferences underexplored. \modelname{} bridges this gap through a context-conditioned Intent Token and a dual-granularity objective, jointly optimizing semantic ID prediction and Intent-Semantic Alignment.

\begin{figure*}[!ht]
\centering
\includegraphics[width=0.95\linewidth]{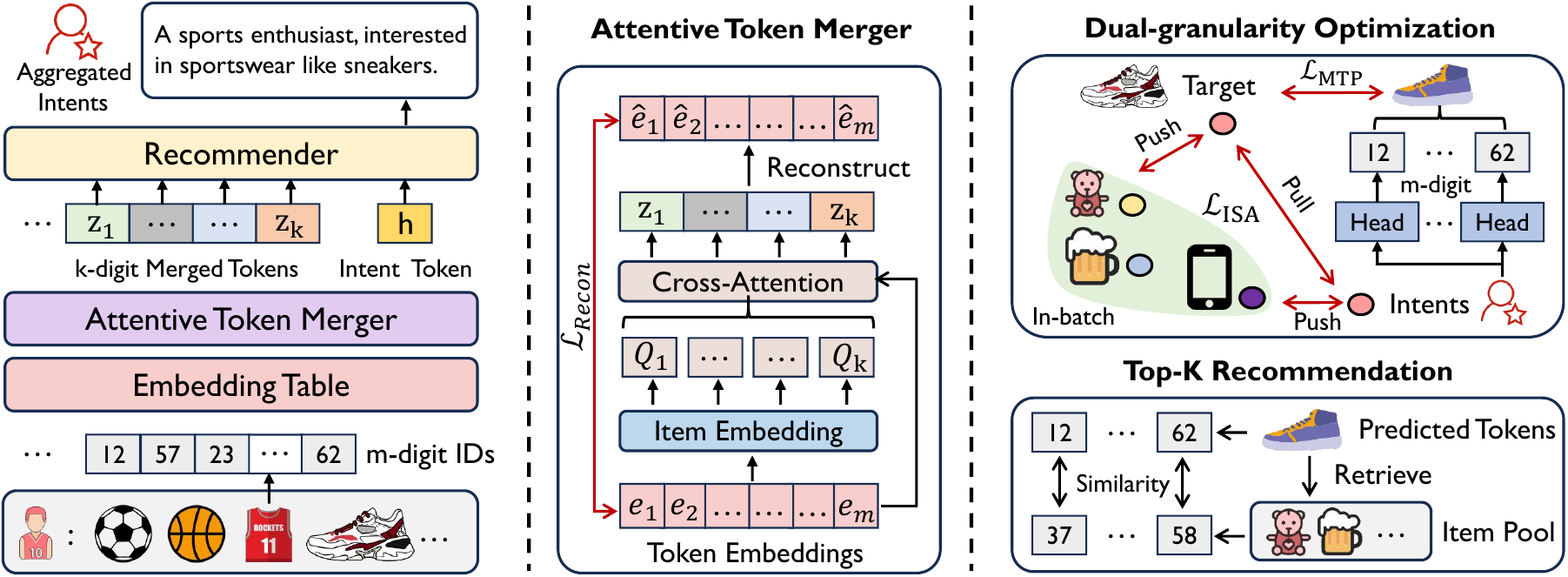}
\caption{Overview of \modelname{}. The sequence encoding flow (left) compresses fine-grained historical item tokens into compact latents via ATM (middle) to feed the sequential recommender. The right panel illustrates the dual-granularity training objectives and the parallel retrieval procedure during inference.} 
\label{Fig: model}
\end{figure*}

\section{Methodology}\label{sec:method}
We illustrate the architecture of  \modelname{} in Figure~\ref{Fig: model}. We first introduce its fine-grained tokenization and adaptive compression, followed by intent-centric sequential modeling and parallel candidate scoring. The complete training and inference procedures are provided in Appendix~\ref{app: code}.

\subsection{Tokenization and Compression} \label{subsec: compress}
\subsubsection{Semantic Tokenization via OPQ}

We employ OPQ~\cite{ge2013optimized,jegou2010product} to decompose item embeddings into $m$ orthogonal subspaces, where each item $i$ is represented by semantic IDs $(c_1,\dots,c_m)$ quantized by subspace-specific codebooks $\{\mathcal{C}^{(k)}\}_{k=1}^{m}$. This factorized representation distributes item information across subspaces and supports parallel token prediction. We adopt long semantic IDs (e.g., $m=32$) to preserve fine-grained item information, and map each code $c_k$ into a learnable embedding $\mathbf{e}_k\in\mathbb{R}^{d}$, forming the representation matrix $\mathbf{E}_i\in\mathbb{R}^{m\times d}$.

\subsubsection{Latent Token Distillation via ATM}
To resolve the conflict between semantic expressiveness and recommendation efficiency, we introduce ATM to adaptively distill long semantic IDs into compact latents.

\noindent\textbf{\textit{Content-adaptive Query Generation.}}
To guide semantic extraction, we aggregate the token sequence into an item-level summary $\mathbf{s}_i$ via a projection $f_s$, and subsequently map it into $k$ content-conditioned queries $\mathbf{Q}_i \in \mathbb{R}^{k \times d}$ via a projector $f_q$:
\begin{equation} \label{eq:q_gen}
\mathbf{s}_i = f_s \left(\mathbf{e_1}, \dots, \mathbf{e_m} \right), \quad \mathbf{Q}_i = f_q (\mathbf{s}_i).
\end{equation}
These content-adaptive queries allow the merger to flexibly concentrate on distinct item properties.

\noindent\textbf{\textit{Attentive Token Merging.}}
To anchor subspace identities, we inject learnable positional embeddings $\mathbf{P} \in \mathbb{R}^{m \times d}$ into the sequence: $\tilde{\mathbf{E}}_i = \mathbf{E}_i + \mathbf{P}$. 
We then use the content-adaptive queries $\mathbf{Q}_i$ to selectively distill the uncompressed semantic sequence $\tilde{\mathbf{E}}_i$ through multi-head cross-attention, producing compact latent tokens $\mathbf{Z}_i \in \mathbb{R}^{k \times d}$ with $k \ll m$:
\begin{equation} \label{eq:atm}
\mathbf{Z}_i = f_{out} \left( f_{attn} \left( \mathbf{Q}_i, \tilde{\mathbf{E}}_i, \tilde{\mathbf{E}}_i \right) \right),
\end{equation}
where $f_{attn}$ performs the scaled dot-product cross-attention and $f_{out}$ contains the multi-layer perceptron (MLP) layers and layer normalization.

\noindent\textbf{\textit{High-Fidelity Token Reconstruction.}}
To preserve fine-grained semantics during compression, we introduce a lightweight upsampling MLP $\text{Decoder}(\cdot)$ that reconstructs the $k$ latent tokens to the original sequence length:
$\hat{\mathbf{E}}_i=\text{Decoder}(\mathbf{Z}_i)\in\mathbb{R}^{m\times d}$.
The reconstruction objective minimizes the mean cosine distance between the reconstructed and original token embeddings across all $m$ subspaces:
\begin{equation} \label{eq:recon_loss}
\mathcal{L}_{\mathrm{Recon}} = 1 - \frac{1}{m} \sum_{r=1}^m \frac{\hat{\mathbf{e}}_{i,r} \cdot \mathbf{e}_{i,r}}{\|\hat{\mathbf{e}}_{i,r}\| \cdot \|\mathbf{e}_{i,r}\|},
\end{equation}
where $\mathbf{e}_{i,r}$ represents the raw embedding of the $r$-th ground-truth semantic token.

\subsection{Intent-Centric Sequential Modeling} \label{subsec: intent_token}

\subsubsection{Modeling Evolving User Intent}
The compressed latent tokens provide fine-grained item semantics, but next-item prediction further requires modeling how user preferences evolve across interactions.
We therefore introduce a context-conditioned \emph{Intent Token} to aggregate relevant item semantics with historical context for next-item prediction.

\noindent\textbf{\textit{Intent Token Formulation.}}
At each step $t$, the Intent Token $\mathbf{h}_t \in \mathbb{R}^d$ is initialized with the item-level summary $\mathbf{s}_t$ from Eq.~\ref{eq:q_gen} and appended to the compressed latent set $\mathbf{Z}_t$:
$\tilde{\mathbf{Z}}_t=[\mathbf{Z}_t;\mathbf{h}_t]\in\mathbb{R}^{(k+1)\times d}$.
The interaction sequence is then constructed as
$\mathbf{H}=[\tilde{\mathbf{Z}}_1,\dots,\tilde{\mathbf{Z}}_L]$.
This design provides each interaction step with a dedicated state that can summarize the current item and accumulate preference information from preceding behaviors, allowing $\mathbf{h}_t$ to represent the user preference up to step $t$.

\noindent\textbf{\textit{Step-wise Causal Attention.}}
We then apply step-wise causal attention over $\mathbf{H}$ to capture preference transitions while preserving intra-item semantic interactions. Tokens at step $t$ cannot attend to future blocks $\tilde{\mathbf{Z}}_{>t}$. Within each step, the latent tokens in $\mathbf{Z}_t$ attend to one another, while $\mathbf{h}_t$ attends to both $\mathbf{Z}_t$ and preceding blocks $\tilde{\mathbf{Z}}_{<t}$. This enables the Intent Token to integrate the current interaction semantics with historical preferences for next-item prediction.

\subsubsection{Dual-Granularity Alignment}
We optimize the intent representation with a dual-granularity objective: fine-grained token prediction for accurate semantic ID generation, and item-level alignment to ensure that the learned preference remains predictive of the target item.

\noindent\textbf{\textit{Fine-grained Token Prediction.}}
We adopt the multi-token prediction (MTP) objective~\cite{gloeckle2024better} to reconstruct the target semantic IDs from the final intent state $\mathbf{h}_i \in \mathbb{R}^d$. We normalize and project $\mathbf{h}_i$ into $m$ independent representations $\{\mathbf{h}^{(k)}_i\}_{k=1}^m$. For each digit $k$, the probability of the $v$-th codeword in the codebook $\mathcal{C}^{(k)}$ is: 
\begin{equation} \label{eq:mtp_prob}
P(c_k = v \mid \mathbf{h}_i) =
\frac{\exp(\mathbf{h}^{(k)\top}_i \mathbf{e}^{(k)}_v / \gamma)}
{\sum_{v'=1}^{M} \exp(\mathbf{h}^{(k)\top}_i \mathbf{e}^{(k)}_{v'} / \gamma)},
\end{equation}
where $\mathbf{e}^{(k)}_v$ is the embedding of the $v$-th codeword in the $k$-th subspace, $M$ is the codebook size, and $\gamma$ is a temperature parameter. The MTP loss is the average negative log-likelihood across all $m$ digits:
\begin{equation} \label{eq:mtp_loss}
\mathcal{L}_{\mathrm{MTP}} = -\frac{1}{m} \sum_{k=1}^m \log P(c_{tgt, k} \mid \mathbf{h}_i),
\end{equation}
where $c_{tgt, k}$ is the target codeword index at the $k$-th position.

\noindent\textbf{\textit{Intent-Semantic Alignment (ISA).}}
While MTP supervises fine-grained semantic ID prediction, it does not explicitly ensure that the learned intent representation remains predictive of the target item as a whole. We therefore introduce Intent-Semantic Alignment (ISA) to align the intent representation $\mathbf{h}_i$ with the holistic representation $\mathbf{s}_{i+1}$ of the target item. Since popular items occur more often as in-batch negatives, we calibrate similarity scores by item popularity~\cite{bengio2008adaptive, yi2019sampling, agarwal2025pinrec}:

\begin{equation}
\phi(\mathbf{h}_{\text{i}}, \mathbf{s}_j) = \text{sim}(\mathbf{h}_{\text{i}}, \mathbf{s}_j) / \tau - b_j,
\end{equation}
where $\tau$ is the temperature, $b_j=\beta \cdot \log(f_j)/\max_{i\in\mathcal{I}}\log(f_i)$ is the popularity bias, $f_j$ is the interaction frequency of item $j$, and $\beta$ controls the calibration strength. The ISA loss for batch $\mathcal{B}$ is:
\begin{equation} \label{eq:isa_loss}
\mathcal{L}_{\mathrm{ISA}} = - \log \frac{\exp(\phi(\mathbf{h}_{\text{i}}, \mathbf{s}_{i+1}))}{\sum_{j \in \mathcal{B}} \exp(\phi(\mathbf{h}_{\text{i}}, \mathbf{s}_j))},
\end{equation}

\noindent\textbf{\textit{Joint Objective.}}
The final training objective integrates the dual-granularity intent alignments with the structural token reconstruction into a unified optimization pipeline:
\begin{equation}
\mathcal{L} = \mathcal{L}_{\mathrm{MTP}} + \lambda \mathcal{L}_{\mathrm{ISA}} + \alpha \mathcal{L}_{\mathrm{Recon}},
\end{equation}
where $\lambda$ and $\alpha$ control the contributions of the item-level alignment and reconstruction objectives, respectively.

\subsection{Efficient Inference via Holistic Scoring} \label{subsec: infer} 
During inference, we employ a holistic candidate scoring strategy~\cite{jegou2010product} to perform exact retrieval over the entire item pool $\mathcal{I}$. Bypassing the heavy candidate-wise dot-products inherent in traditional dense retrieval~\cite{kang2018self, sun2019bert4rec}, our approach decouples probability computation from item scoring via two efficient vectorized steps:

\noindent\textbf{Parallel Subspace Matching.} 
We project the aggregated intent state $\mathbf{h}_i$ into $m$ subspace queries $\{\mathbf{h}_i^{(1)}, \dots, \mathbf{h}_i^{(m)}\}$ to compute log-probability distributions over all codebooks in parallel.  The entry $\mathbf{P}[k, v]$ representing the log-probability of the $v$-th codeword in the $k$-th subspace is derived as:

\begin{equation} \label{eq:sim}
\mathbf{P}[k, v] = \log \frac{\exp(\mathbf{h}_i^{(k)} \cdot \mathbf{e}^{(k)}_v / \gamma)}{\sum_{v'=1}^{M} \exp(\mathbf{h}_i^{(k)} \cdot \mathbf{e}^{(k)}_{v'} / \gamma)},
\end{equation}
where $\gamma$ is the prediction temperature. Crucially, this operation's complexity scales only with the vocabulary footprint ($m \times M$), remaining entirely independent of the candidate pool size $|\mathcal{I}|$.

\noindent\textbf{Vectorized Score Gathering.} 
Since each candidate item $j \in \mathcal{I}$ is pre-quantized as a discrete token tuple $\mathbf{c}_j = (c_{j,1}, \dots, c_{j,m})$, we eliminate online embedding computations entirely. Instead, candidate indices are utilized to directly retrieve and aggregate individual log-probabilities from the matrix $\mathbf{P}$ via a fast vectorized gather-and-sum pipeline:

\begin{small}
\begin{equation} \label{eq:score}
\text{Score}(j) = \sum_{k=1}^{m} \mathbf{P}[k, c_{j,k}].
\end{equation}
\end{small}
This design shifts runtime candidate ranking into a high-speed memory lookup-and-sum process, ensuring minimal retrieval latency.

\begin{table*}[!ht]
\tablestyle{1.8pt}{1.05}
\centering
\begin{tabular}{@{} l *{12}{>{\centering\arraybackslash}p{0.068\textwidth}} @{}}
  \toprule
  \multicolumn{1}{c}{\multirow{2.5}{*}{\textbf{Model}}} & \multicolumn{4}{c}{\textbf{Sports}} & \multicolumn{4}{c}{\textbf{Beauty}} & \multicolumn{4}{c}{\textbf{Toys}} \\
  \cmidrule(lr){2-5} \cmidrule(lr){6-9}\cmidrule(lr){10-13}
  & \textbf{R@5} & \textbf{N@5} & \textbf{R@10} & \textbf{N@10} & \textbf{R@5} & \textbf{N@5} & \textbf{R@10} & \textbf{N@10} & \textbf{R@5} & \textbf{N@5} & \textbf{R@10} & \textbf{N@10} \\
  \midrule
  \multicolumn{13}{@{}c}{\textit{Item ID-based}} \\
  \midrule
HGN & 0.0189 & 0.0120 & 0.0313 & 0.0159 & 0.0325 & 0.0206 & 0.0512 & 0.0266 & 0.0321 & 0.0221 & 0.0497 & 0.0277 \\
SASRec & 0.0233 & 0.0154 & 0.0350 & 0.0192 & 0.0387 & 0.0249 & 0.0605 & 0.0318 & 0.0463 & 0.0306 & 0.0675 & 0.0374 \\
S$^3$-Rec & 0.0251 & 0.0161 & 0.0385 & 0.0204 & 0.0387 & 0.0244 & 0.0647 & 0.0327 & 0.0443 & 0.0294 & 0.0700 & 0.0376 \\
ICLRec & 0.0274 & 0.0179 & 0.0426 & 0.0228 & 0.0488 & 0.0322 & 0.0739 & 0.0403 & 0.0573 & 0.0395 & 0.0812 & 0.0472 \\
ELCRec & 0.0276 & 0.0180 & 0.0414 & 0.0225 & 0.0502 & 0.0351 & 0.0727 & 0.0423 & \underline{0.0580} & \underline{0.0398} & \underline{0.0815} & \underline{0.0474} \\

\midrule
\multicolumn{13}{@{}c}{\textit{Semantic ID-based}} \\
\midrule

TIGER & 0.0264 & 0.0181 & 0.0400 & 0.0225 & 0.0454 & 0.0321 & 0.0648 & 0.0384 & 0.0521 & 0.0371 & 0.0712 & 0.0432 \\
ETEGRec & 0.0248 & 0.0159 & 0.0404 & 0.0209 & 0.0420 & 0.0271 & 0.0668 & 0.0350 & 0.0445 & 0.0286 & 0.0716 & 0.0373 \\
ActionPiece & 0.0285 & 0.0187 & \underline{0.0441} & 0.0238 & 0.0484 & 0.0322 & \underline{0.0762} & 0.0412 & 0.0473 & 0.0310 & 0.0725 & 0.0391 \\
RPG & \underline{0.0292} & \underline{0.0196} & 0.0435 & \underline{0.0242} & \underline{0.0512} & \underline{0.0355} & 0.0757 & \underline{0.0436} & 0.0557 & 0.0384 & 0.0777 & 0.0454 \\

\midrule
\textbf{\modelname{}} & \textbf{0.0341} & \textbf{0.0238} & \textbf{0.0494} & \textbf{0.0287} & \textbf{0.0610} & \textbf{0.0433} & \textbf{0.0856} & \textbf{0.0513} & \textbf{0.0698} & \textbf{0.0514} & \textbf{0.0940} & \textbf{0.0591} \\

\textbf{Improv.} & +16.78\% & +21.43\% & +12.02\% & +18.60\% & +19.14\% & +21.97\% & +12.34\% & +17.66\% & +20.34\% & +29.15\% & +15.34\% & +24.68\% \\
\bottomrule
\end{tabular}
\caption{Performance comparison of \modelname{} and baselines on three benchmark datasets. The best and second-best results are denoted in \textbf{bold} and \underline{underline}, respectively. The ``Improv.'' row indicates the percentage improvement of \modelname{} over the strongest baseline. All results are averaged over five runs with different random seeds.}
\label{tab:overall}
\end{table*}

\begin{table*}[!h]
\tablestyle{1.8pt}{1.05}
\centering
\begin{tabular}{@{} l *{12}{>{\centering\arraybackslash}p{0.068\textwidth}} @{}}
  \toprule
  \multicolumn{1}{c}{\multirow{2.5}{*}{\textbf{Model}}} & \multicolumn{4}{c}{\textbf{CDs}} & \multicolumn{4}{c}{\textbf{Baby}} & \multicolumn{4}{c}{\makebox[0pt]{\textbf{Pets}}} \\ 
  \cmidrule(lr){2-5} \cmidrule(lr){6-9} \cmidrule(lr){10-13}
  & \textbf{R@5} & \textbf{N@5} & \textbf{R@10} & \textbf{N@10} & \textbf{R@5} & \textbf{N@5} & \textbf{R@10} & \textbf{N@10} & \textbf{R@5} & \textbf{N@5} & \textbf{R@10} & \textbf{N@10} \\
  \midrule
  \multicolumn{13}{@{}c}{\textit{Item ID-based}} \\
  \midrule
HGN & 0.0259 & 0.0153 & 0.0467 & 0.0220 & 0.0159 & 0.0122 & 0.0278 & 0.0167 & 0.0270 & 0.0206 & 0.0446 & 0.0272 \\
SASRec & 0.0351 & 0.0177 & 0.0619 & 0.0263 & 0.0114 & 0.0058 & 0.0212 & 0.0089 & 0.0237 & 0.0127 & 0.0414 & 0.0184 \\
S$^3$-Rec & 0.0213 & 0.0130 & 0.0375 & 0.0182 & 0.0246 & 0.0159 & 0.0410 & 0.0211 & 0.0385 & 0.0254 & 0.0585 & 0.0318 \\
ICLRec & 0.0477 & 0.0312 & 0.0733 & 0.0394 & 0.0250 & 0.0163 & 0.0401 & 0.0211 & 0.0431 & 0.0292 & 0.0651 & 0.0363 \\
ELCRec & 0.0455 & 0.0299 & 0.0697 & 0.0376 & 0.0247 & 0.0157 & 0.0406 & 0.0209 & 0.0407 & 0.0272 & 0.0623 & 0.0341 \\

\midrule
\multicolumn{13}{@{}c}{\textit{Semantic ID-based}} \\
\midrule

TIGER & \underline{0.0492} & {0.0329} & \underline{0.0748} & \underline{0.0411} & 0.0160 & 0.0101 & 0.0260 & 0.0133 & 0.0291 & 0.0188 & 0.0473 & 0.0247 \\
ETEGRec & 0.0426 & 0.0278 & 0.0673 & 0.0357 & 0.0138 & 0.0084 & 0.0246 & 0.0118 & 0.0372 & 0.0232 & 0.0604 & 0.0306 \\
ActionPiece & 0.0481 & 0.0312 & 0.0738 & 0.0395 & \underline{0.0269} & \underline{0.0175} & \underline{0.0424} & \underline{0.0225} & 0.0410 & 0.0270 & 0.0639 & 0.0343 \\
RPG & 0.0488 & \underline{0.0330} & 0.0708 & 0.0400 & 0.0254 & 0.0174 & 0.0394 & 0.0220 & \underline{0.0448} & \underline{0.0302} & \underline{0.0674} & \underline{0.0374} \\

\midrule
\textbf{\modelname{}} & \textbf{0.0547} & \textbf{0.0370} & \textbf{0.0800} & \textbf{0.0451} & \textbf{0.0308} & \textbf{0.0214} & \textbf{0.0462} & \textbf{0.0263} & \textbf{0.0494} & \textbf{0.0338} & \textbf{0.0706} & \textbf{0.0407} \\

\textbf{Improv.} & +11.18\% & +12.46\% & +6.95\% & +9.73\% & +14.50\% & +22.29\% & +8.96\% & +16.89\% & +10.27\% & +11.92\% & +4.75\% & +8.82\% \\
\bottomrule
\end{tabular}
\caption{Performance comparison on the CDs, Baby and Pets datasets. The notations follow Table~\ref{tab:overall}.}
\label{tab:overall_2}
\end{table*}

\section{Experiment}\label{sec:experiment}

In this section, we conduct extensive experiments on nine real-world datasets to answer the following research questions:

\begin{itemize}
[leftmargin=*]
    \item \textbf{RQ1:} How does \modelname{} perform compared with strong discriminative and generative recommendation baselines?
    \item \textbf{RQ2:} Does \modelname{} effectively decouple tokenization granularity from recommendation efficiency by outperforming direct modeling with short semantic IDs (e.g., 4 tokens) through learned compression of long semantic IDs (e.g., 32 tokens)?
    \item \textbf{RQ3:} How do \modelname{}'s components contribute to overall performance?
\end{itemize}

\subsection{Experimental Setup}
\fakeparagraph{Datasets}
We evaluate \modelname{} across nine benchmark domains from two generations of the Amazon Reviews collection~\cite{mcauley2015image, hou2024bridging}: Sports, Beauty, Toys, CDs, Office, Pets, and Baby from the 2014 edition~\cite{mcauley2015image}, alongside Science and Instruments from the 2023 version~\cite{hou2024bridging}. Following standard generative protocols~\cite{rajput2023recommender, hou2023learning, hou2025generating}, sequences are chronologically sorted and split via a leave-last-out strategy (the last interaction for testing, the second-to-last for validation, and the rest for training).  Dataset statistics are provided in Appendix~\ref{app: dataset}.


\fakeparagraph{Baselines}
We compare \modelname{} against five item ID–based methods (HGN~\cite{ma2019hierarchical}, SASRec~\cite{kang2018self}, $\textbf{S}^3$-Rec~\cite{zhou2020s3}, ICLRec~\cite{chen2022intent}, ELCRec~\cite{liu2024end}) and four semantic ID–based generative pipelines (TIGER~\cite{rajput2023recommender}, ActionPiece~\cite{hou2025actionpiece}, ETEGRec~\cite{liu2025generative}, RPG~\cite{hou2025generating}). Detailed descriptions of all baselines are provided in Appendix~\ref{app: baseline}.

\fakeparagraph{Evaluation Metrics and Implementation Details}
All models are evaluated under full candidate ranking using Recall@$K$ and NDCG@$K$ ($K \in \{5, 10\}$). Implementation details are provided in Appendix~\ref{app: implementation}.

\subsection{Overall Performance}
Tables~\ref{tab:overall} and~\ref{tab:overall_2} report the results on six representative datasets, with the remaining results provided in Appendix~\ref{app:additional_overall}. \modelname{} consistently achieves the best performance across all datasets, with average relative improvements of 7.49\% in Recall@10 and 12.92\% in NDCG@10 over the strongest baselines. The gains are especially pronounced on Sports, Beauty, and Toys, where NDCG@10 improves by 18.60\%, 17.66\%, and 24.68\%, respectively, while remaining stable on the substantially larger CDs dataset.

The results further highlight the importance of both components in \modelname{}. Compared with RPG, which similarly represents items using long OPQ IDs but compresses them through mean pooling, \modelname{} achieves clear gains across all datasets. This comparison isolates the benefit of adaptive compression: rather than uniformly collapsing all subspace tokens, ATM selectively preserves the semantic information most useful for downstream recommendation. Meanwhile, strong item ID-based intent models such as ELCRec can outperform previous semantic ID-based methods on datasets such as Toys, indicating that richer item representations alone do not guarantee better recommendation. By combining fine-grained long-ID semantics with a context-conditioned Intent Token and item-level alignment, \modelname{} more effectively transforms item semantics into predictive user intent, leading to consistent improvements over both generative and discriminative baselines.


\subsection{Analysis of Tokenization}
A core premise of \modelname{} is that decoupling tokenization granularity from recommendation efficiency is essential. In this section, we investigate this hypothesis by analyzing scalability with respect to semantic ID length and comparing our learned compression against direct short tokenization.

\begin{figure}[!h]
    \centering
    \includegraphics[width=\linewidth]{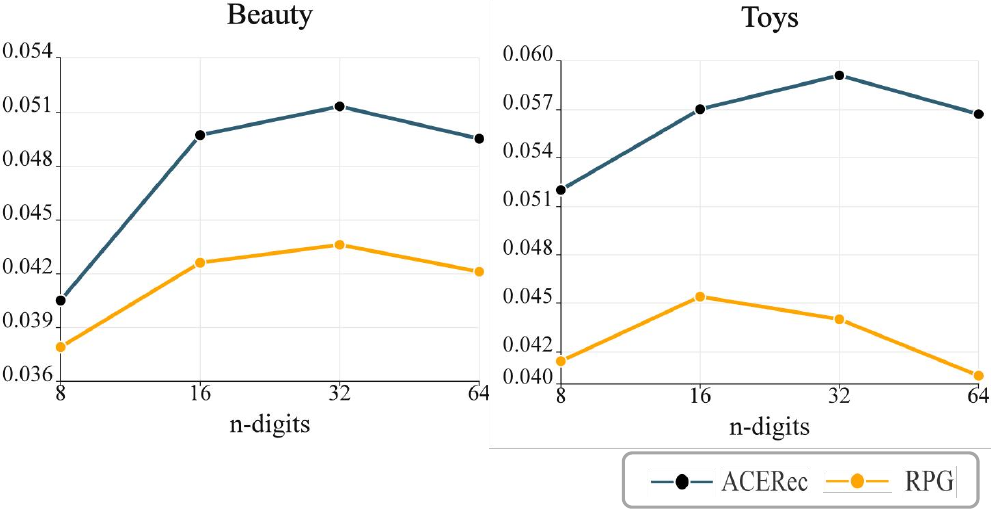}
    
    \caption{NDCG@10 performance under different semantic ID lengths $m$.}
    \label{Fig: ablation_digits}
\end{figure}

\fakeparagraph{Scalability of Semantic ID Lengths}
The semantic ID length $m$ governs the granularity of item representation. To analyze its impact, we vary $m \in \{8, 16, 32, 64\}$ while fixing the compression ratio at $r=8$. Figure~\ref{Fig: ablation_digits} compares the NDCG@10 performance of \modelname{} and RPG on Beauty and Toys.

\modelname{} consistently outperforms RPG across all tested ID lengths on both datasets. Its performance improves substantially from $m=8$ to $m=16$, reaches the best result at $m=32$, and remains competitive at $m=64$. In comparison, RPG exhibits weaker scaling behavior: its performance declines beyond $m=32$ on Beauty and beyond $m=16$ on Toys. These results indicate that ATM and $\mathcal{L}_{\text{Recon}}$ enable \modelname{} to utilize longer semantic IDs more effectively, while excessively long IDs eventually provide diminishing returns. We therefore adopt $m=32$ as the default semantic ID length. Appendix~\ref{app:compress} further examines sensitivity to the compression ratio, while Appendix~\ref{app:resolution_capacity} analyzes the interaction between semantic ID length $m$ and latent capacity $k$.

\begin{figure}[!h]
    \centering
    \includegraphics[width=\linewidth]{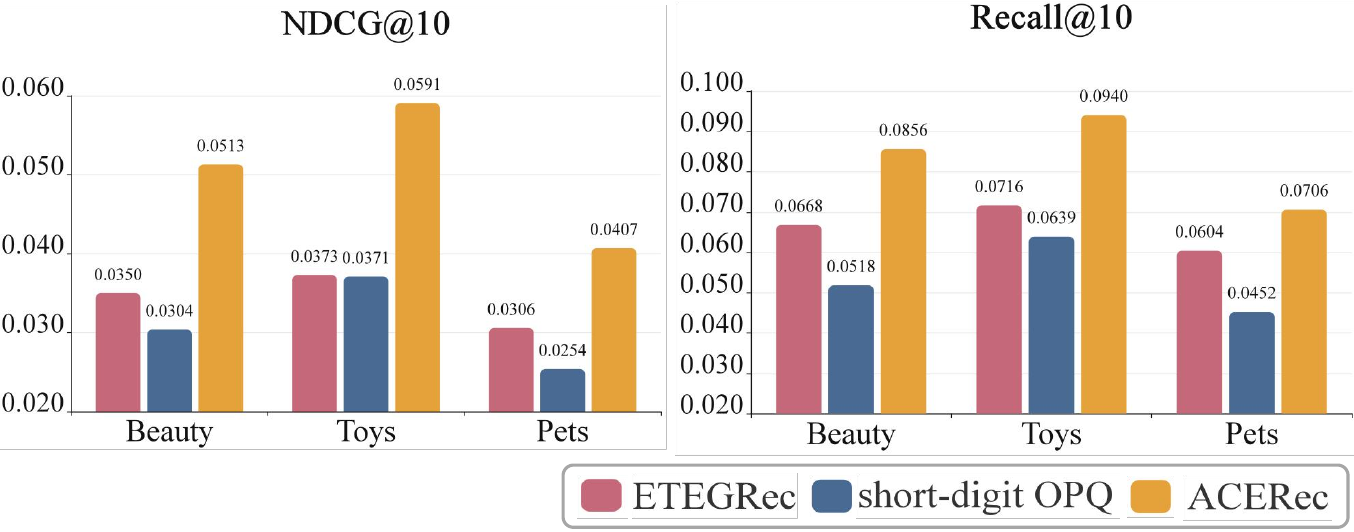}
    
    \caption{Performance comparison between \modelname{} and short-digit RQ / OPQ baselines. All models use the same input length for the recommender.}
    \label{Fig: ablation_short_OPQ}
\end{figure}

\fakeparagraph{Effectiveness of Decoupling Strategy} To verify whether the gains originate from preserving the expressive potential of long semantic IDs, we compare \modelname{} with two short-ID alternatives while fixing the recommender input to four tokens per item. Short-digit OPQ directly quantizes each item into only four independent semantic codes, providing a compact but limited representation. ETEGRec~\cite{liu2025generative} adopts RQ-based semantic IDs, consisting of three residual semantic tokens and one conflict token. In contrast, \modelname{} first constructs a 32-token OPQ semantic ID to capture complementary fine-grained semantics, and then compresses it into four latent tokens through ATM. As shown in Figure~\ref{Fig: ablation_short_OPQ}, \modelname{} consistently achieves the best Recall@10 and NDCG@10 across all datasets.

The comparison reveals two observations. First, under the same short-ID budget, RQ-based ETEGRec outperforms Short-digit OPQ by 24.88\% Recall@10 and 12.05\% NDCG@10 on average, suggesting that residual quantization is more effective than shallow OPQ when the code length is highly constrained. Second, over short-digit OPQ, \modelname{} achieves gains of 56.18\% in Recall@10 and 62.76\% in NDCG@10. Against ETEGRec, the corresponding gains are 25.44\% and 46.01\%, respectively. Unlike directly quantizing into short IDs, \modelname{} preserves richer semantics through long OPQ representations and selectively distills them into compact latent tokens, demonstrating the advantage of decoupling tokenization granularity from sequential modeling.

\subsection{Ablation Study of \modelname{}}
We conduct a comprehensive ablation study consisting of two parts: \textit{Component Contributions of \modelname{}} and \textit{Fine-grained Design Analysis}.

\fakeparagraph{Impact of Key Components}
As shown in Table~\ref{tab:ablation_ACERec}, we progressively evolve the baseline framework to isolate the gains of each core module.

\begin{itemize}[leftmargin=*]
    \item \textbf{Efficacy of Adaptive Compression:} We investigate the compression mechanism by replacing naive rigid pooling with ATM. It yields consistent improvements over the baseline: specifically, NDCG@10 increases from 0.0454 to 0.0470 on Toys and from 0.0220 to 0.0226 on Baby, demonstrating that attentive merging effectively preserves fine-grained semantics.
    
    \item \textbf{Impact of Intent-Centric Predictive Anchor:} Introducing the dedicated Intent Token triggers a substantial performance surge, particularly in ranking precision. For instance, NDCG@10 increases sharply by 19.57\% (0.0470 $\to$ 0.0562) on the Toys domain, highlighting the necessity of maintaining a context-conditioned proactive state to dynamically query historical latents rather than relying on standard passive token sequences.
    
    \item \textbf{Impact of Training Objectives:} 
    Adding $\mathcal{L}_{\text{ISA}}$ (Row 4) improves Toys NDCG@10 to 0.0579 by aligning user intent with target-item semantics. Further introducing $\mathcal{L}_{\text{Recon}}$ (Row 5) achieves the best results, confirming that reconstruction helps preserve fine-grained semantics during compression.
    The sensitivity analysis of the loss weights is provided in Appendix~\ref{app:sensitivity}.
\end{itemize}

\begin{table}[!h]
\tablestyle{3.5pt}{1.1}
\centering
{
\begin{tabular}{@{}l cc cc@{}}
  \toprule
  \multirow{2.5}{*}{\textbf{Model Variant}} & \multicolumn{2}{c}{\textbf{Toys}} & \multicolumn{2}{c}{\textbf{Baby}} \\
  \cmidrule(lr){2-3} \cmidrule(l){4-5}
  & \textbf{R@10} & \textbf{N@10} & \textbf{R@10} & \textbf{N@10} \\
  \midrule
  Base (RPG) & 0.0777 & 0.0454 & 0.0394 & 0.0220 \\
  + ATM  & 0.0802 & 0.0470 & 0.0406 & 0.0226 \\
  + Intent Token             & 0.0869 & 0.0562 & 0.0415 & 0.0242 \\
  + $\mathcal{L}_{\text{ISA}}$ & 0.0916 & 0.0579 & 0.0427 & 0.0249 \\
  + $\mathcal{L}_{\text{Recon}}$ (\modelname{}) & \textbf{0.0940} & \textbf{0.0591} & \textbf{0.0462} & \textbf{0.0263} \\
  \bottomrule
\end{tabular}}
\caption{Ablation analysis of \modelname{} components.}
\label{tab:ablation_ACERec}
\end{table}

\fakeparagraph{Ablation of Architectural Designs}
Beyond validating the overall framework, we further examine whether our specific architectural choices are necessary. We therefore conduct fine-grained ablations on (i) the \textit{merging strategy} and (ii) the \textit{generation anchor}, as reported in Table~\ref{tab:ablation_meger}.

\begin{itemize}[leftmargin=*]
    \item \textbf{Impact of Merging Strategy.}
    We replace ATM with MLP, mean pooling, and convolution while keeping the remaining components unchanged. ATM consistently performs best across datasets and metrics, while convolution remains competitive in ranking precision but lags in recall. These results highlight the importance of the merging strategy: fixed aggregation may dilute informative subspace features, whereas ATM uses cross-attention to adaptively distill content-dependent semantics into compact latents. Appendix~\ref{app:atm_visualization} provides a complementary qualitative view of this content-adaptive behavior.
    \item \textbf{Impact of Generation Anchor Strategy.} 
    We compare the Intent Token with three static prediction anchors derived from the last item’s latent set $\mathbf{Z}_L$: \textit{Last-Token}, \textit{Mean-Pooling}, and \textit{MLP}. As shown in Table~\ref{tab:ablation_meger} (lower), the Intent Token consistently performs best, improving Recall@10 from 0.0805 to 0.0869 and NDCG@10 from 0.0496 to 0.0562 on Toys. Unlike static aggregation, the Intent Token serves as a context-conditioned prediction anchor that is dynamically updated through step-wise causal attention, allowing it to integrate relevant semantics from historical interactions and better represent evolving user intent.
\end{itemize}

\begin{table}[h!]
\tablestyle{5.0pt}{1.05}
\centering
{
\begin{tabular}{@{}lcccc@{}}
  \toprule
  \multicolumn{1}{c}{\multirow{2.5}{*}{\textbf{Model}}} & \multicolumn{2}{c}{\textbf{Toys}} & \multicolumn{2}{c}{\textbf{Baby}} \\
  \cmidrule(lr){2-3} \cmidrule(l){4-5}
  & \textbf{R@10} & \textbf{N@10} & \textbf{R@10} & \textbf{N@10} \\
\midrule
\multicolumn{5}{@{}c}{\textit{Ablation of Merging Strategies}} \\
\midrule
MLP & 0.0869 & 0.0518 & 0.0433 & 0.0236 \\
Mean-Pooling & 0.0890 & 0.0563 & 0.0428 & 0.0256 \\
Convolution & 0.0891 & 0.0570 & 0.0443 & 0.0259 \\
ATM & \textbf{0.0940} & \textbf{0.0591} & \textbf{0.0462} & \textbf{0.0263} \\
  \midrule
  \multicolumn{5}{@{}c}{\textit{Ablation of Generation Anchor}} \\
\midrule
Last-Token & 0.0802 & 0.0470 & 0.0406 & 0.0226 \\
Mean-Pooling & 0.0805 & 0.0496 & \textbf{0.0420} & 0.0227 \\
MLP & 0.0771 & 0.0466 & 0.0393 & 0.0214 \\
Intent-Token & \textbf{0.0869} & \textbf{0.0562} & {0.0415} & \textbf{0.0242} \\
\bottomrule
\end{tabular}}
\caption{Fine-grained ablation of architectural designs.}
\label{tab:ablation_meger}
\end{table}

\subsection{Further Analysis} \label{subsubsec: cold_start}
\begin{figure}[!h]
    \centering
    \includegraphics[width=\linewidth]{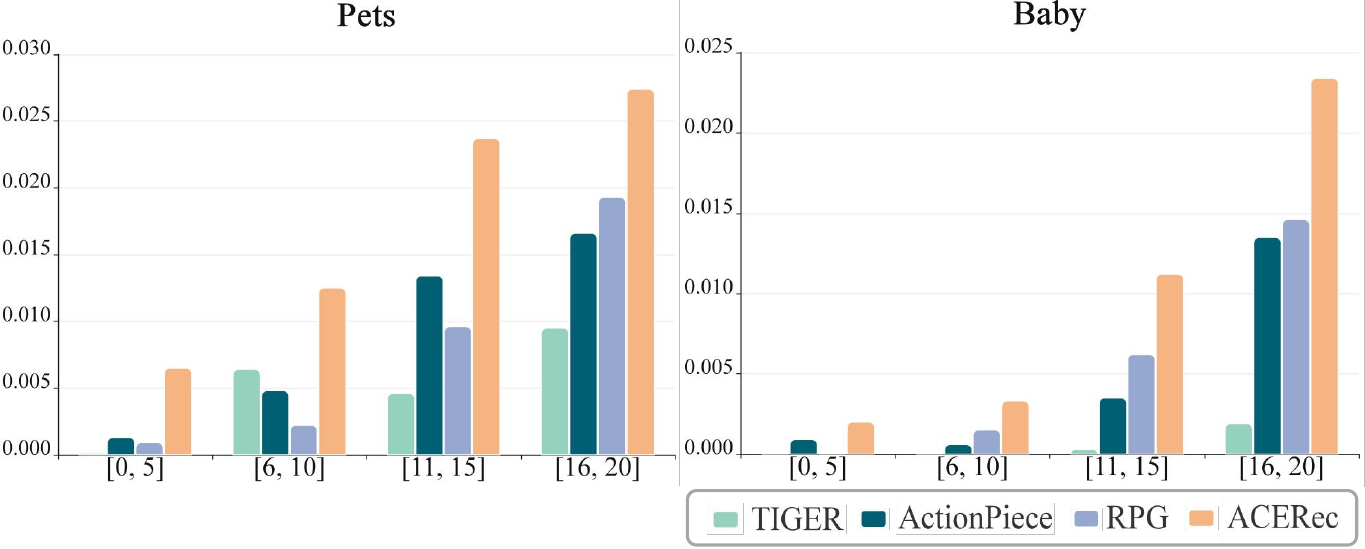}
    
    \caption{Cold-start analysis on Pets and Baby datasets, with NDCG@10 as the evaluation metric.} 
    \label{Fig: cold}
\end{figure}

\fakeparagraph{Robustness to Cold-Start Items} 
To evaluate the robustness of \modelname{} to extreme data sparsity, we partition the test items into four buckets based on their interaction frequencies in the training set: $[0, 5]$, $[6, 10]$, $[11, 15]$, and $[16, 20]$. As shown in Figure~\ref{Fig: cold}, \modelname{} consistently outperforms all baselines across every frequency interval, maintaining a substantial lead even as interaction sparsity varies.

In the extreme-sparsity bucket (\eg $[0, 5]$), TIGER and RPG degrade to near-zero performance, as they struggle to learn transferable semantic representations from items with few interactions. In contrast, \modelname{} remains competitive, demonstrating stronger generalization through its effective use of item semantics under sparse interaction signals. Moreover, as item frequency increases (\eg $[16, 20]$), \modelname{} further widens the gap, demonstrating stable gains beyond the coldest regime. 
This robustness may stem from ACERec's effective use of item semantics: ATM preserves informative features from long semantic IDs, while ISA improves the learned user intent by aligning it with target-item semantics, benefiting recommendation under sparse interactions.
Additional cold-start results are provided in Appendix~\ref{app: cold_start}.

\begin{figure}[!h]
    \centering
    \includegraphics[width=\linewidth]{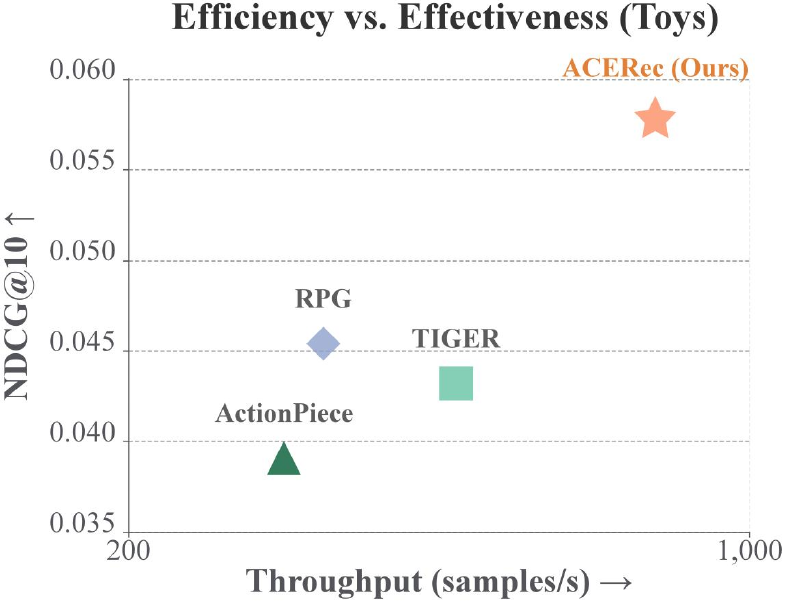}
    
    \caption{Inference Efficiency and Recommendation Performance on the Toys Dataset.}
    
    \label{Fig: efficiency}
\end{figure}

\fakeparagraph{Inference Efficiency vs. Effectiveness}
We evaluate the trade-off between inference throughput (samples/s) and NDCG@10 on the Toys dataset (\Cref{Fig: efficiency}). While autoregressive models like ActionPiece and TIGER are hindered by costly beam search, and RPG is bottlenecked by iterative graph decoding, \modelname{} achieves an average speedup of $2.2 \times$ over them. Notably, when TIGER is configured with the same number of tokens (e.g., 4) for the recommender, \modelname{} achieves higher inference throughput due to its parallel decoding nature, while simultaneously attaining the best overall performance. These advantages in speed and ranking are directly attributed to our ATM-based decoupling mechanism, which successfully exploits fine-grained semantic sequences through an efficient parallel framework, effectively bridging the gap between representational richness and inference throughput. 
We further provide the complexity comparison in Appendix~\ref{app:complexity} and the analysis of training convergence in Appendix~\ref{app:convergence}.

\section{Conclusion}\label{sec:conclusion}

In this work, we presented \modelname{} to address the trade-off between expressive semantic representation and efficient sequential modeling in generative recommendation. \modelname{} decouples tokenization granularity from recommendation complexity by using ATM to adaptively distill long semantic IDs into compact latent tokens, together with a reconstruction objective that preserves fine-grained semantics during compression. Built on these compact representations, a context-conditioned Intent Token serves as a dynamic prediction anchor for modeling evolving user preferences, while a dual-granularity objective jointly optimizes semantic ID prediction and Intent-Semantic Alignment. Extensive experiments on nine real-world benchmarks demonstrate that \modelname{} consistently outperforms strong discriminative and generative baselines, achieving an average relative improvement of 12.92\% in NDCG@10 over the strongest competitors. Further analysis shows that \modelname{} remains effective for low-frequency items, indicating improved robustness under sparse interaction signals.


\bibliography{aaai2027}
\twocolumn[{
	\renewcommand\twocolumn[1][]{#1}
	\begin{center}
		\textbf{\Large Appendix:\\Unleash the Potential of Long Semantic IDs for Generative Recommendation}\end{center}
}]

\appendix

\section{Experimental Setting Details}
\subsection{Task Formulation} \label{app: task}
Let $\mathcal U$ and $\mathcal I$ denote the set of users and items, respectively. For each user $u \in \mathcal U$, the interaction history is represented as a chronological sequence $\mathcal S_{u} {=} \{i_1, i_2, \dots, i_L\}$. The goal of sequential recommendation is to predict the next item $i_{L+1}$ from the candidate set $\mathcal I$~\cite{kang2018self, sun2019bert4rec}. This can be formulated as learning a conditional distribution $P(i_{L+1}\mid \mathcal S_{u})$. 

In the paradigm of Generative Sequential Recommendation, each item $i$ is represented not by an atomic ID, but by $m$ discrete tokens (i.e., semantic IDs), denoted as $\mathbf{c}_i = (c_{i, 1}, c_{i, 2}, \dots, c_{i, m})$. Consequently, the recommendation task is reformulated as generating the token sequence of the target item given the user's history~\cite{rajput2023recommender, hou2025generating}. Two decoding strategies exist for this generation process: (1) \textit{Autoregressive Decoding}: Early approaches~\cite{rajput2023recommender, liu2025generative} flatten the history into a long stream and generate tokens step-by-step via beam search~\cite{hokamp2017lexically}, modeling the conditional probability $P(c_{i, k}\mid c_{i, <k},\mathcal{S}_u)$. (2) \textit{Parallel Decoding}: Recent methods~\cite{hou2025generating, hou2023learning} utilize OPQ~\cite{ge2013optimized, jegou2010product}, which decomposes item embeddings into orthogonal subspaces. This structure supports a conditional independence assumption, enabling the parallel prediction of all $m$ tokens, i.e.  $P(\mathbf{c}_{i_{L+1}}\mid\mathcal{S}_u)\approx\prod_{k=1}^mP(c_{{i_{L+1}}, k}\mid\mathcal{S}_u)$.

\subsection{Baseline Details} \label{app: baseline}
We compare \modelname{} with nine state-of-the-art baselines. These methods are categorized into two groups: (1) \textbf{item ID-based methods}, which rely on unique item indices, optionally enhanced with auxiliary features, and (2) \textbf{semantic ID-based methods}, which leverage item content (e.g., brand and price). Detailed descriptions of these baselines are as follows:

\begin{itemize}
[leftmargin=*]
    \item \textbf{HGN}~\cite{ma2019hierarchical}:  utilizes a hierarchical gating network to capture both long-term and short-term user interests from interaction sequences.
    \item \textbf{SASRec}~\cite{kang2018self}: employs a self-attentive Transformer decoder to model sequential dynamics and predict the next item.
    \item $\textbf{S}^3$\textbf{-Rec}~\cite{zhou2020s3}: utilizes mutual information maximization to pre-train the model with correlations among items, attributes, and subsequences.
    \item \textbf{ICLRec}~\cite{chen2022intent}: learns latent user intents via clustering and contrastive learning within an Expectation-Maximization framework.
    \item \textbf{ELCRec}~\cite{liu2024end}: unifies behavior representation and clustering into an end-to-end framework to efficiently capture latent user intents.
    \item \textbf{TIGER}~\cite{rajput2023recommender}: quantizes items into hierarchical semantic IDs via RQ-VAE and autoregressively generates the next item's tokens.
    \item \textbf{ActionPiece}~\cite{hou2025actionpiece}:  proposes context-aware tokenization by representing actions as feature sets and merging co-occurring patterns.
    \item \textbf{ETEGRec}~\cite{liu2025generative}: jointly optimizes the item tokenizer and the generative recommender using multi-view alignment objectives.
    \item \textbf{RPG}~\cite{hou2025generating}: utilizes OPQ to construct long semantic IDs and adopts an MTP objective for parallel generation.
\end{itemize}

\subsection{Dataset Statistics} \label{app: dataset}
The experiments are conducted on nine public benchmark datasets derived from two distinct generations of the Amazon product reviews collection~\cite{mcauley2015image, hou2024bridging}. Specifically, we incorporate seven domains from the 2014 edition: \textbf{Sports}, \textbf{Beauty}, \textbf{Toys}, \textbf{CDs}, \textbf{Office}, \textbf{Pets}, and \textbf{Baby}, as well as two categories from the 2023 edition: \textbf{Science} and \textbf{Instruments}. The 2014 subsets comprise user reviews and item metadata collected between May 1996 and July 2014~\cite{mcauley2015image}, whereas the 2023 subsets capture contemporary e-commerce interaction dynamics up to 2023~\cite{hou2024bridging}.

The data have been processed following standard evaluation protocols for sequential recommendation: item sequences for each user are sorted chronologically by timestamps, and users with fewer than five interactions are excluded to ensure sufficient context. During the training phase, the length of user historical sequences is truncated to the 50 most recent interactions to maintain computational efficiency. The statistics of the processed datasets are summarized in Table~\ref{tab:dataset}.

\begin{table}[h] %
  \small
  \centering
\tablestyle{3.5pt}{1.1} 
\begin{tabular}{ccccc}
  \toprule
  \textbf{Datasets} & \textbf{\#Users} & \textbf{\#Items} & \textbf{\#Interactions} & \textbf{Avg.~$t$}\\
  \midrule
  \textbf{Sports}  & 18,357            & 35,598           & 260,739            & 8.32 \\
  \textbf{Beauty}  & 22,363            & 12,101           & 176,139            & 8.87 \\
  \textbf{Toys}    & 19,412            & 11,924           & 148,185            & 8.63 \\
  \textbf{CDs}    & 75,258           & 64,443           & 1,022,334            & 14.58 \\
  \textbf{Baby}    & 19,446            & 7,051           & 160,792            & 8.27 \\
  \textbf{Pets}    & 19,856 & 8,510 & 157,836 & 7.95 \\
  \textbf{Office}    & 4,906            & 2,421           & 53,258            & 10.86 \\
  \textbf{Science} & 50,986 & 25,849 & 412,947 & 8.10 \\
    \textbf{Instruments}     & 57,440 & 24,588 & 511,836 & 8.91 \\
  \bottomrule
\end{tabular}
\caption{Statistics of the processed datasets. ``Avg.~$t$'' denotes the average number of interactions per input sequence.}
\label{tab:dataset}
\end{table}


\subsection{Implementation Details} \label{app: implementation}
\noindent\textbf{Baselines.} To ensure a rigorous comparison, we adopt the reported results from Rajput \textit{et al.}~\cite{rajput2023recommender} for HGN, SASRec, $\textrm{S}^3$-Rec, and TIGER on the Sports, Beauty, and Toys datasets. For all other datasets and baselines, we reproduce the results using their official implementations or the RecBole library~\cite{zhao2021recbole}, carefully tuning hyperparameters as suggested in their original papers. 

\noindent\textbf{Unified Semantic Encoder.} To ensure a fair comparison, we consistently employ \texttt{sentence-t5-base}~\cite{ni2022sentence} as the universal semantic encoder for \textit{all} semantic ID-based methods, ensuring that any performance gains are attributed to our proposed architecture rather than superior raw semantic embeddings. We use the FAISS library~\cite{douze2025faiss} to implement OPQ.

\noindent\textbf{\modelname{}.} We implement our method using HuggingFace transformers~\cite{wolf2020transformers} and align the model configuration with RPG~\cite{hou2025generating} for fair comparison. Specifically, we utilize a 2-layer Transformer~\cite{vaswani2017attention} decoder with an embedding dimension of $d{=}448$, a feed-forward network dimension of 1024, and 4 attention heads. Regarding the semantic ID length $m$, we focus on $m=32$ to fully leverage the expressive power of long sequences, while the default compression ratio is 8. The subspace codebook vocabulary size $M$ is configured as 256 for the Amazon Reviews 2014 datasets and 512 for the 2023 edition. Key hyperparameters are configured as follows: popularity debiasing scaling coefficient $\beta=0.02$, MTP generation temperature $\gamma=0.03$, and ISA contrastive temperature $\tau=0.07$.

All experiments were conducted on a single NVIDIA RTX 6000 Ada GPU (48GB). To manage the search space efficiently, we adopt a \textit{two-stage tuning protocol}. First, we fix the ISA weight $\lambda=0.1$, the reconstruction loss weight $\alpha=0.03$ and perform a grid search over the learning rate in $\{0.003, 0.005\}$ and batch size in $\{64, 256\}$, totaling 4 hyperparameter settings. Second, based on the optimal configuration from the first stage, we jointly fine-tune $\lambda$ within $\{0.1, 0.15, 0.2\}$ and $\alpha$ within $\{0.03, 0.05\}$. This results in a total of 9 experimental runs per dataset. For inference, unlike generative baselines such as ActionPiece and RPG that inherently rely on beam search~\cite{rajput2023recommender} or cumbersome decoding parameter tuning~\cite{hou2025generating}, \modelname{} offers a highly concise retrieval paradigm requiring zero inference hyperparameter tuning; we simply load the best checkpoint based on validation NDCG@10 performance and perform holistic candidate scoring.

\begin{algorithm}[t]
\caption{Training Procedure of \modelname{}}
\label{alg: training}
\begin{algorithmic}[1]
\Require 
    User sequences $\mathcal{D} = \{(\mathcal{S}_u, i_{tgt})\}$, OPQ codebooks $\{\mathcal{C}^{(k)}\}$.
\Ensure Trained Model parameters $\Theta$.

\State \textbf{Initialize} parameters $\Theta$ randomly.

\While{not converged}
    \For{each sample $(\mathcal{S}_u, i_{tgt})$ in $\mathcal{D}$}
        \State \textcolor{gray}{// 1. Token Compression}
        \For{each item $i_t \in \mathcal{S}_u$}
            \State Generate queries $\mathbf{Q}_t$ and intent $\mathbf{h}_t$ by \Cref{eq:q_gen}.
            \State Merge latents $\mathbf{Z}_t$ via \Cref{eq:atm}.
            \State Set block $\tilde{\mathbf{Z}}_t = [\mathbf{Z}_t; \mathbf{h}_t]$.
        \EndFor
        \State Extract target latents $\mathbf{Z}_{tgt}$.
        \State Compute $\mathcal{L}_{\text{Recon}}$ via \Cref{eq:recon_loss}.
        
        \State \textcolor{gray}{// 2. Sequence modeling}
        \State Construct Step-wise Causal Mask $\mathbf{M}$.
        \State Aggregate Intent via Recommender.
        
        \State \textcolor{gray}{// 3. Dual-Granularity Optimization}
        \State Compute $\mathcal{L}_{\text{MTP}}$ by \Cref{eq:mtp_loss}.
        \State Compute $\mathcal{L}_{\text{ISA}}$ by \Cref{eq:isa_loss}.
        \State $\mathcal{L} \leftarrow \mathcal{L}_{\text{MTP}} + \lambda \mathcal{L}_{\text{ISA}} + \alpha \mathcal{L}_{\text{Recon}}$.
        \State Update $\Theta \leftarrow \Theta - \eta \nabla \mathcal{L}$.
    \EndFor
\EndWhile
\end{algorithmic}
\end{algorithm}

\begin{algorithm}[ht]
\caption{Inference via Holistic Candidate Scoring}
\label{alg: inference}
\begin{algorithmic}[1]
\Require 
    Trained model $\Theta$, User history $\mathcal{S}_u$, Item set $\mathcal{I}$.
\Ensure Top-$K$ recommendations.

\State \textcolor{gray}{// 1. Contextualized Encoding}
\For{each item $i_t \in \mathcal{S}_u$}
    \State Generate latents $\mathbf{Z}_t$ and intent $\mathbf{h}_t$ using Equations ~(\ref{eq:q_gen}) and~(\ref{eq:atm}).
    \State Set block $\tilde{\mathbf{Z}}_t = [\mathbf{Z}_t; \mathbf{h}_t]$.
\EndFor
\State Construct Step-wise Causal Mask $\mathbf{M}$.
\State Aggregate Intent via Recommender.

\State \textcolor{gray}{// 2. Parallel Subspace Matching}
\State Project $\mathbf{h}$ into $m$ queries $\{\mathbf{h}^{(1)}, \dots, \mathbf{h}^{(m)}\}$.
\For{digit $k = 1$ to $m$ \textbf{in parallel}}
    \State Compute logits $\mathbf{P}[k, \cdot]$ by~\Cref{eq:sim}.
\EndFor

\State \textcolor{gray}{// 3. Vectorized Score Gathering}
\State \textbf{Gather} scores $\mathbf{S} \in \mathbb{R}^{|\mathcal{I}|}$ by~\Cref{eq:score}.

\State \textbf{return} Top-$K$ indices from $\mathbf{S}$.
\end{algorithmic}
\end{algorithm}

\section{Procedure and Discussion of \modelname{}} 
\subsection{Overall Procedure of \modelname{}} \label{app: code}
In this section, we provide the detailed algorithmic procedures for both the training and inference phases of \modelname{}. Algorithm~\ref{alg: training} outlines the end-to-end training flow, systematically integrating fine-grained tokenization via OPQ and attentive token merging, latent sequence modeling with step-wise causal masking, and the dual-granularity optimization strategy. Algorithm~\ref{alg: inference} presents the inference phase, utilizing our holistic candidate scoring strategy. Unlike autoregressive approaches, it demonstrates how \modelname{} achieves efficient exact retrieval by computing subspace probabilities in parallel and performing vectorized score aggregation over the entire item catalog.

\subsection{Discussion of \modelname{}} \label{app: discussion}
In this section, we provide an in-depth discussion on the connections and distinctions between \modelname{} and existing generative recommendation paradigms, followed by a formal complexity analysis.

\fakeparagraph{Comparison with Generative Baselines}
\modelname{} evolves from existing generative recommenders by explicitly resolving the ``Granularity Mismatch'' dilemma through decoupled architecture designs.

\begin{itemize}
[leftmargin=*]
    \item \textbf{vs. TIGER}: Unlike TIGER~\cite{rajput2023recommender}, which couples code length with inference latency due to the auto-regressive decoding dependencies, \modelname{} decouples them by leveraging long, attribute-rich IDs ($m \ge 32$) for expressive item representation while compressing them into compact latents ($k=4$) to maintain computational tractability.
    \item \textbf{vs. RPG}: Compared to RPG~\cite{hou2025generating}, which suffers from semantic blurring due to rigid mean pooling, \modelname{} adaptively filters and distills high-density information based on item context, preserving critical intra-item structures. 
\end{itemize}

\begin{table*}[!h]
\tablestyle{1.8pt}{1.05}
\centering
\begin{tabular}{@{} l *{12}{>{\centering\arraybackslash}p{0.068\textwidth}} @{}}
  \toprule
  \multicolumn{1}{c}{\multirow{2.5}{*}{\textbf{Model}}} & \multicolumn{4}{c}{\textbf{Office}} & \multicolumn{4}{c}{\textbf{Science}} & \multicolumn{4}{c}{\makebox[0pt]{\textbf{Instruments}}} \\ 
  \cmidrule(lr){2-5} \cmidrule(lr){6-9} \cmidrule(lr){10-13}
  & \textbf{R@5} & \textbf{N@5} & \textbf{R@10} & \textbf{N@10} & \textbf{R@5} & \textbf{N@5} & \textbf{R@10} & \textbf{N@10} & \textbf{R@5} & \textbf{N@5} & \textbf{R@10} & \textbf{N@10} \\
  \midrule
  \multicolumn{13}{@{}c}{\textit{Item ID-based}} \\
  \midrule
HGN & 0.0342 & 0.0303 & 0.0580 & 0.0389 & 0.0164 & 0.0131 & 0.0265 & 0.0169 & 0.0212 & 0.0162 & 0.0351 & 0.0215 \\
SASRec & 0.0493 & 0.0255 & 0.0881 & 0.0379 & 0.0097 & 0.0052 & 0.0182 & 0.0080 & 0.0085 & 0.0044 & 0.0174 & 0.0073 \\
S$^3$-Rec & \underline{0.0587} & \underline{0.0381} & \underline{0.0989} & \underline{0.0509} & 0.0254 & 0.0167 & 0.0402 & 0.0215 & 0.0337 & 0.0219 & 0.0538 & 0.0283 \\
ICLRec & 0.0516 & 0.0342 & 0.0822 & 0.0440 & 0.0269 & 0.0179 & 0.0403 & 0.0222 & 0.0349 & 0.0230 & 0.0545 & 0.0292 \\
ELCRec & 0.0481 & 0.0157 & 0.0728 & 0.0403 & 0.0265 & 0.0177 & 0.0393 & 0.0219 & 0.0354 & 0.0236 & 0.0535 & 0.0294 \\

\midrule
\multicolumn{13}{@{}c}{\textit{Semantic ID-based}} \\
\midrule

TIGER & 0.0359 & 0.0224 & 0.0607 & 0.0303 & 0.0234 & 0.0150 & 0.0356 & 0.0190 & 0.0337 & 0.0219 & 0.0513 & 0.0276 \\
ETEGRec & 0.0336 & 0.0216 & 0.0612 & 0.0304 & 0.0270 & 0.0173 & 0.0426 & 0.0224 & \underline{0.0379} & 0.0246 & \underline{0.0579} & \underline{0.0311} \\
ActionPiece & 0.0512 & 0.0334 & 0.0858 & 0.0444 & 0.0267 & 0.0172 & 0.0422 & 0.0222 & 0.0374 & 0.0235 & 0.0477 & 0.0304 \\
RPG & 0.0579 & {0.0381} & 0.0917 & 0.0490 & \underline{0.0287} & \underline{0.0189} & \underline{0.0429} & \underline{0.0234} & 0.0375 & \underline{0.0248} & 0.0557 & 0.0307 \\

\midrule
\textbf{\modelname{}} & \textbf{0.0673} & \textbf{0.0456} & \textbf{0.1036} & \textbf{0.0572} & \textbf{0.0295} & \textbf{0.0198} & \textbf{0.0439} & \textbf{0.0244} & \textbf{0.0388} & \textbf{0.0259} & \textbf{0.0579} & \textbf{0.0321} \\

\textbf{Improv.} & +14.65\% & +19.69\% & +4.75\% & +12.38\% & +2.79\% & +4.76\% & +2.33\% & +4.27\% & +2.37\% & +4.44\% & +0.00\% & +3.22\% \\
\bottomrule
\end{tabular}
\caption{Performance comparison on the Office, Science and Instruments datasets. The notations follow Table~\ref{tab:overall}.}
\label{tab:overall_3}
\end{table*}

    

\fakeparagraph{Complexity and Efficiency} \label{app:complexity}
Let $L$ be the sequence length, $d$ the dimension. TIGER flattens all $m$ tokens per item, incurring $O(L^2n^2d)$ self-attention complexity, which is prohibitive for expressive IDs. RPG aggregates tokens into a single vector, reducing complexity to $O(L^2d)$ at the cost of semantic truncation. \modelname{} introduces a linear cost of $O(Lmkd)$ for ATM distillation, while the Transformer backbone operates on $N = L {\times} (k+1)$ tokens, resulting in $O(L^2 k^2 d)$ complexity.

Since $k \ll m$, \modelname{} achieves a favorable trade-off: \modelname{} has significantly lower complexity than directly modeling the raw sequence ($k^2 \ll m^2$) while retaining fine-grained semantics. Regarding inference retrieval, \modelname{} avoids TIGER's serial token-generation bottleneck and eliminates the cumbersome multi-parameter generation tuning required by ActionPiece and RPG. By performing exact candidate matching in a fully parallel manner via vectorized matrix gather operations, \modelname{} achieves high evaluation throughput without inference hyperparameter tuning.

\section{Additional Experimental Results} \label{app: exp}
\subsection{Additional Overall Performance} \label{app:additional_overall}
Table \ref{tab:overall_3} illustrates the overall performance across the three supplementary datasets: Office, Science and Instruments. The experimental results demonstrate that \modelname{} consistently achieves the best results across all evaluation metrics, providing strong evidence of its exceptional generalization ability across different product domains and across different versions of the dataset (spanning the classic Amazon 2014 version and the latest 2023 version). Notably, on the two newer datasets, Science and Instruments, semantic ID-based methods generally demonstrated greater stability than most item ID models, while \modelname{} further raised the performance ceiling.

\subsection{Impact of Compression Ratio} \label{app:compress}
\begin{figure}[!h]
    \centering
    \includegraphics[width=\linewidth]{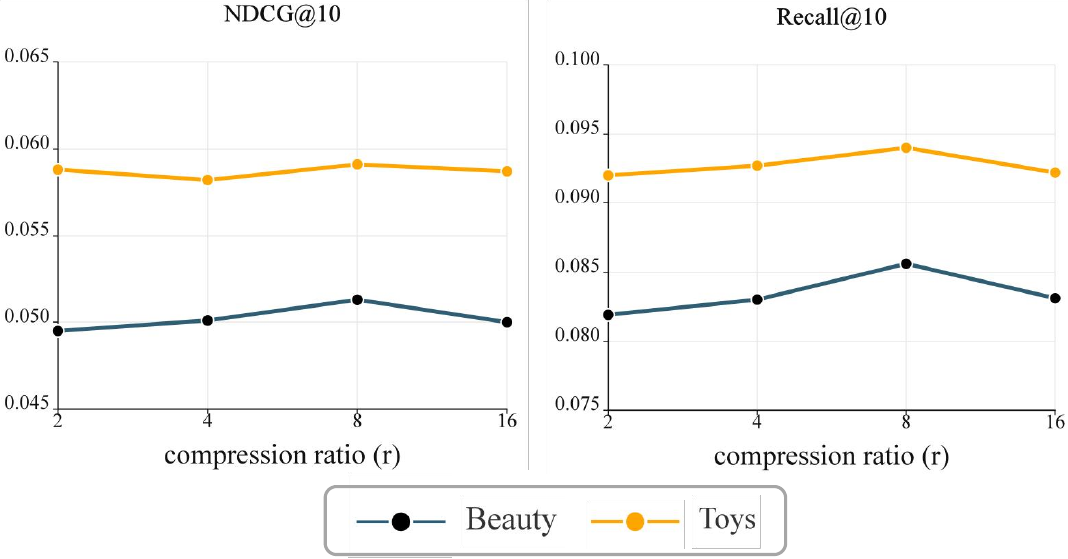}
    \caption{Impact of compression ratio $r$ on recommendation performance.}
    \label{Fig: ablation_compress}
\end{figure}

We investigate the compression ratio $r=m/k$ by varying it within $\{2,4,8,16\}$. As illustrated in Figure~\ref{Fig: ablation_compress}, recommendation performance remains stable across the tested settings and reaches its best overall trade-off at $r=8$.

Lower compression ratios ($r=2,4$) provide only marginal gains from retaining more latent tokens, whereas \modelname{} remains competitive even under the more aggressive setting $r=16$. This stability suggests that ATM can compress long semantic IDs into a small latent set without a substantial loss in recommendation performance. We therefore use $r=8$ as the default compression ratio to balance effectiveness and efficiency.

\subsection{Sensitivity of objective weights} \label{app:sensitivity}
\begin{figure}[!h]
    \centering
    \includegraphics[width=\linewidth]{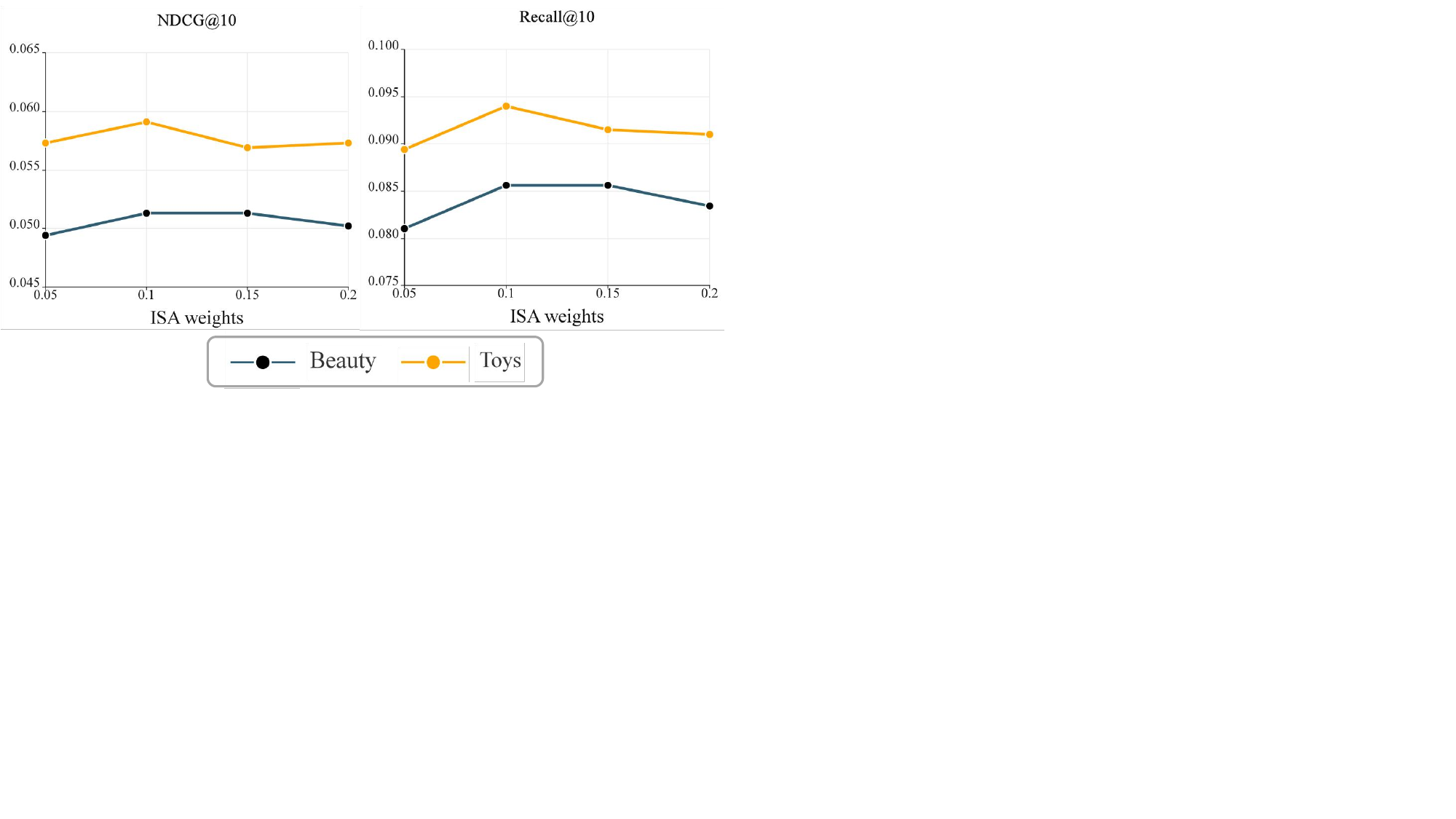}
    
    \caption{Performance comparison of different ISA loss coefficients.}
    \label{Fig: ablation_lambda}
\end{figure}

\fakeparagraph{Sensitivity of ISA Weight $\lambda$}
To analyze how the strength of $\mathcal L_{\mathrm{ISA}}$ affects performance, we vary the loss weight $\lambda$ within the range $\{0.05, 0.1, 0.15, 0.2\}$ on the Beauty and Toys datasets. As illustrated in Figure~\ref{Fig: ablation_lambda}, \modelname{} exhibits remarkable performance robustness to variations in $\lambda$, with overall fluctuations across both NDCG@10 and Recall@10 remaining stable within a tight margin. Crucially, $\lambda = 0.1$ consistently emerges as the optimal choice for both benchmarks. Below this threshold ($\lambda = 0.05$), the collaborative intent modeling is slightly under-regularized; above it ($\lambda \ge 0.15$), the objective leans too heavily toward holistic alignment, marginally distracting from the token-level predictive exactness. This smooth peak validates that our dual-granularity framework acts as a stable and reliable joint objective, and we therefore fix $\lambda = 0.1$ as the default weight and finetune it within $\{0.1, 0.15, 0.2\}$.

\begin{figure}[t]
    \centering
    \includegraphics[width=\linewidth]{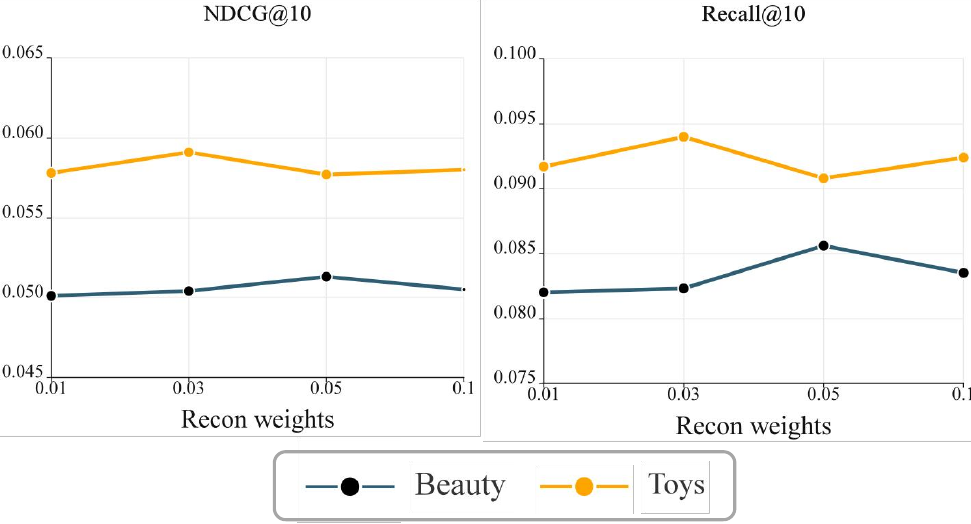}
    
    \caption{Performance comparison of different Reconstruction loss coefficients.}
    \label{Fig: ablation_alpha}
\end{figure}

\fakeparagraph{Sensitivity of Reconstruction Weight $\alpha$}
To analyze how the strength of the continuous token reconstruction regularization affects recommendation performance, we vary the loss weight $\alpha$ within the range $\{0.01, 0.03, 0.05, 0.1\}$ on the Beauty and Toys datasets. As illustrated in Figure~\ref{Fig: ablation_alpha}, \modelname{} exhibits a highly robust performance profile to variations in $\alpha$, with metrics maintaining a stable and elevated plateau across both benchmarks. Specifically, the optimal performance bounds gracefully settle within the $\alpha \in [0.03, 0.05]$ interval, where the Toys dataset peaks at $\alpha = 0.03$ and the Beauty dataset peaks at $\alpha = 0.05$. Below this sweet spot ($\alpha = 0.01$), the geometric constraint safeguarding structural integrity is slightly under-regularized, risking minor attribute information decay during compression; above it ($\alpha = 0.1$), the joint optimization objective over-prioritizes low-level token alignment, marginally distracting the backbone from capturing sequential intent transitions. We therefore fine-tune $\alpha$ within $\{0.03, 0.05\}$.

\subsection{Additional Cold-Start Analysis}\label{app: cold_start}

\begin{figure*}[!ht]
\centering
\includegraphics[width=\linewidth]{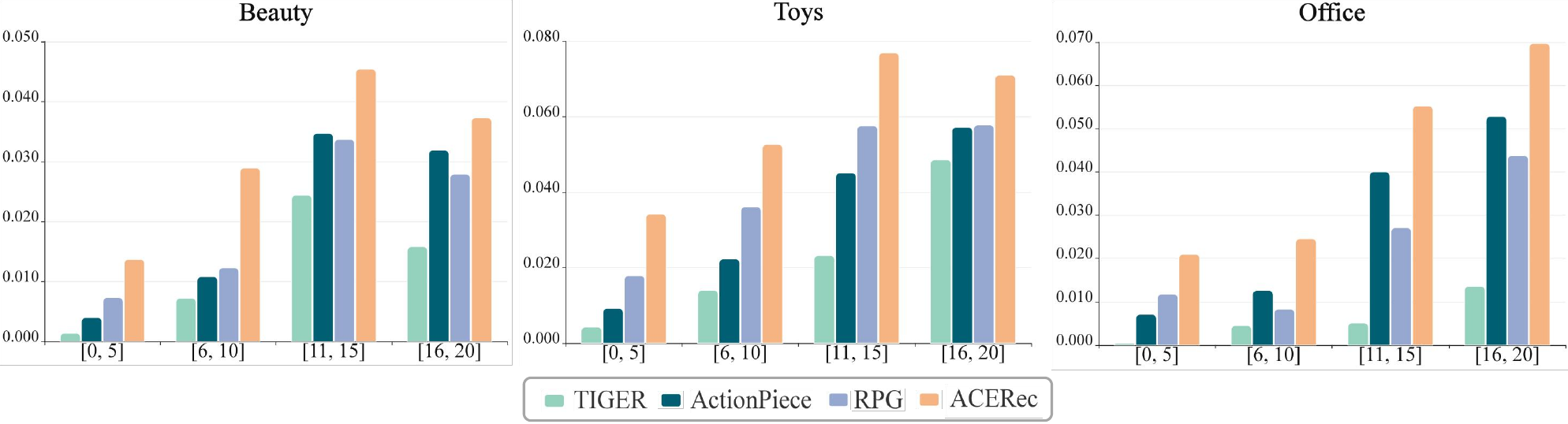}
\caption{Additional cold-start performance (NDCG@10) on Beauty, Toys, and Office datasets.} 
\label{Fig: app_cold}
\end{figure*}

To further substantiate the robustness of \modelname{}, we present a detailed breakdown of item frequency distributions in Table~\ref{tab:cold_start_dist} and extend the cold-start analysis to the Beauty, Toys, and Office datasets (Figure~\ref{Fig: app_cold}) with the same partition.

\noindent\textbf{Prevalence of Long-Tail Items.} The statistics in Table~\ref{tab:cold_start_dist} reveal a pervasive long-tail distribution across all domains. Notably, items in the extreme sparsity bucket ($[0, 5]$) constitute a substantial portion of the test evaluations, ranging from 19.30\% (Baby) to as high as 32.2\% (Toys). When combined with the $[6, 10]$ bucket, tail items account for nearly half of the entire catalog in most datasets. This underscores that resolving low-frequency item recommendation is not merely an edge case but a dominant factor determining overall generative recommendation effectiveness.

\begin{table}[!h]
\centering
\tablestyle{5.0pt}{1.1} 
\begin{tabular}{@{} l ccccc @{}}
\toprule
\multirow{2}{*}{\textbf{Dataset}} & \multicolumn{5}{c}{\textbf{Item Frequency Buckets (\%)}} \\
\cmidrule(l){2-6}
 & $[0, 5]$ & $[6, 10]$ & $[11, 15]$ & $[16, 20]$ & $[21, \infty)$ \\
\midrule
\textbf{Sports}      & 28.0 & 19.1 & 10.7 & 7.3 & 34.9 \\
\textbf{Beauty}      & 27.7 & 19.1 & 11.4 & 6.8 & 35.0 \\
\textbf{Toys}        & 32.2 & 20.6 & 11.5 & 7.5 & 28.2 \\
\textbf{CDs} & 17.5 & 17.1 & 10.0 & 7.6 & 47.8 \\
\textbf{Office}      & 28.4 & 15.1 & 9.4  & 6.8 & 40.3 \\
\textbf{Baby}        & 19.3 & 15.1 & 10.3 & 8.1 & 47.2 \\
\textbf{Science}        & 22.9 & 18.5 & 11.0 & 7.2 & 40.4 \\
\textbf{Instruments}        & 15.1 & 13.9 & 9.2 & 7.0 & 54.8 \\
\bottomrule
\end{tabular}
\caption{Distribution of test items based on their interaction frequency in the training set. The values indicate the percentage (\%) of test items falling into each sparsity bucket.}
\label{tab:cold_start_dist}
\end{table}

\noindent\textbf{Performance Consistency.} As shown in Figure~\ref{Fig: app_cold}, \modelname{} exhibits consistent robustness on three additional datasets, aligning with the observations in the main text. Crucially, in the high-prevalence $[0, 5]$ bucket, \modelname{} significantly outperforms baselines—nearly doubling RPG's performance on Beauty and Office. Given that 27.71\% of Beauty and 28.38\% of Office items fall into this category, this gain translates to a meaningful real-world impact. This strongly confirms that our model design effectively prevents model collapse and enables stable knowledge transfer.

\subsection{Interplay between Semantic Resolution and Latent Capacity} \label{app:resolution_capacity}

\begin{figure}[!h]
\centering
\includegraphics[width=\linewidth]{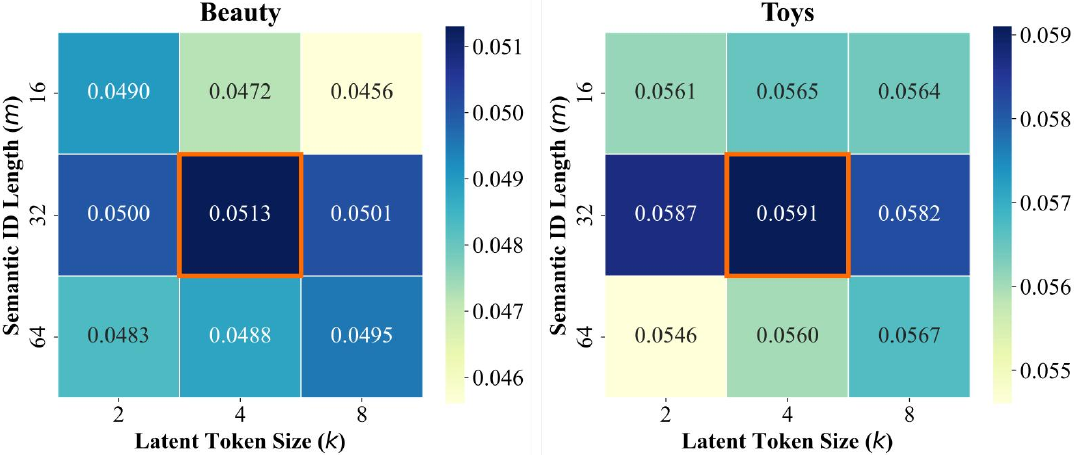}
\caption{NDCG@10 performance under different semantic ID lengths $m$ and compressed latent sizes $k$.} 
\label{Fig: m_k_mat}
\end{figure}

Figure~\ref{Fig: m_k_mat} illustrates the joint performance impact of varying the input semantic ID length $m$ and the compressed latent size $k$ on the Beauty and Toys datasets. Crucially, we observe that the raw input resolution ($m$) clearly dominates the latent token capacity ($k$). For instance, the default configuration ($m=32, k=4$) consistently and significantly outperforms ($m=16, k=8$) across both benchmarks. This empirical pattern confirms that a structural bottleneck in raw semantic fidelity cannot be salvaged by simply increasing the number of compressed latent tokens; a fine-grained tokenized input is an indispensable prerequisite for high-precision generative recommendation. 

Furthermore, the heatmaps reveal that both domains share an identical structural ``sweet spot'' at exactly ($m=32, k=4$), highlighted by the orange borders. This shared optimum strongly justifies the efficacy of our framework: when compressed by our ATM, a highly compact latent space of $k=4$ tokens is entirely sufficient to retain the distilled essence of $32$ raw semantic attributes without unrecoverable information loss. Conversely, further extending the codebooks to an aggressive resolution of $m=64$ leads to performance saturation or minor decline. This indicates that overly elongated codebooks might inject subtle digit-level noise or over-complicate parallel prediction heads. Consequently, we fix ($m=32, k=4$) as the standard architecture across our primary evaluation pipeline to strike a favorable balance between high-fidelity expressiveness and computational efficiency.

\subsection{Comparison of Convergence Rate} \label{app:convergence}

\begin{figure}[!h]
    \centering
    \includegraphics[width=0.95\linewidth]{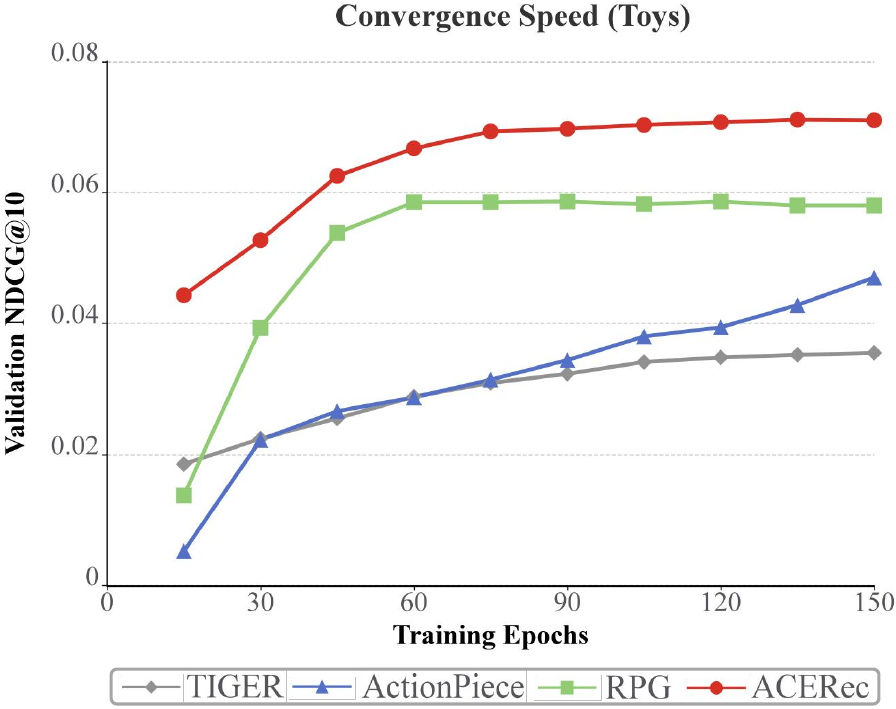}
    
    \caption{Training dynamics and convergence rate on the Toys dataset.}
    
    \label{Fig: convergence}
\end{figure}

We further investigate the learning efficiency and training stability of \modelname{} by tracking the validation NDCG@10 over 150 training epochs on Toys. As shown in Figure~\ref{Fig: convergence}, \modelname{} converges substantially faster and reaches a higher plateau than other generative baselines: its NDCG@10 surpasses 0.06 within only 40 epochs and consistently stabilizes around 0.071 after 70 epochs. In contrast, RPG improves rapidly at the beginning but saturates early at approximately 0.058, potentially due to the limited expressive power of its mean-pooling architecture. Meanwhile, ActionPiece and TIGER exhibit much slower progress and remain below 0.046 and 0.035 even after 150 epochs, reflecting the optimization difficulties. Overall, the training trajectory indicates that \modelname{} is more sample-efficient, as reflected by its faster rise, and achieves a higher performance ceiling, suggesting more effective optimization over long semantic-ID representations.

\begin{figure*}[t]
    \centering
    \includegraphics[width=\linewidth]{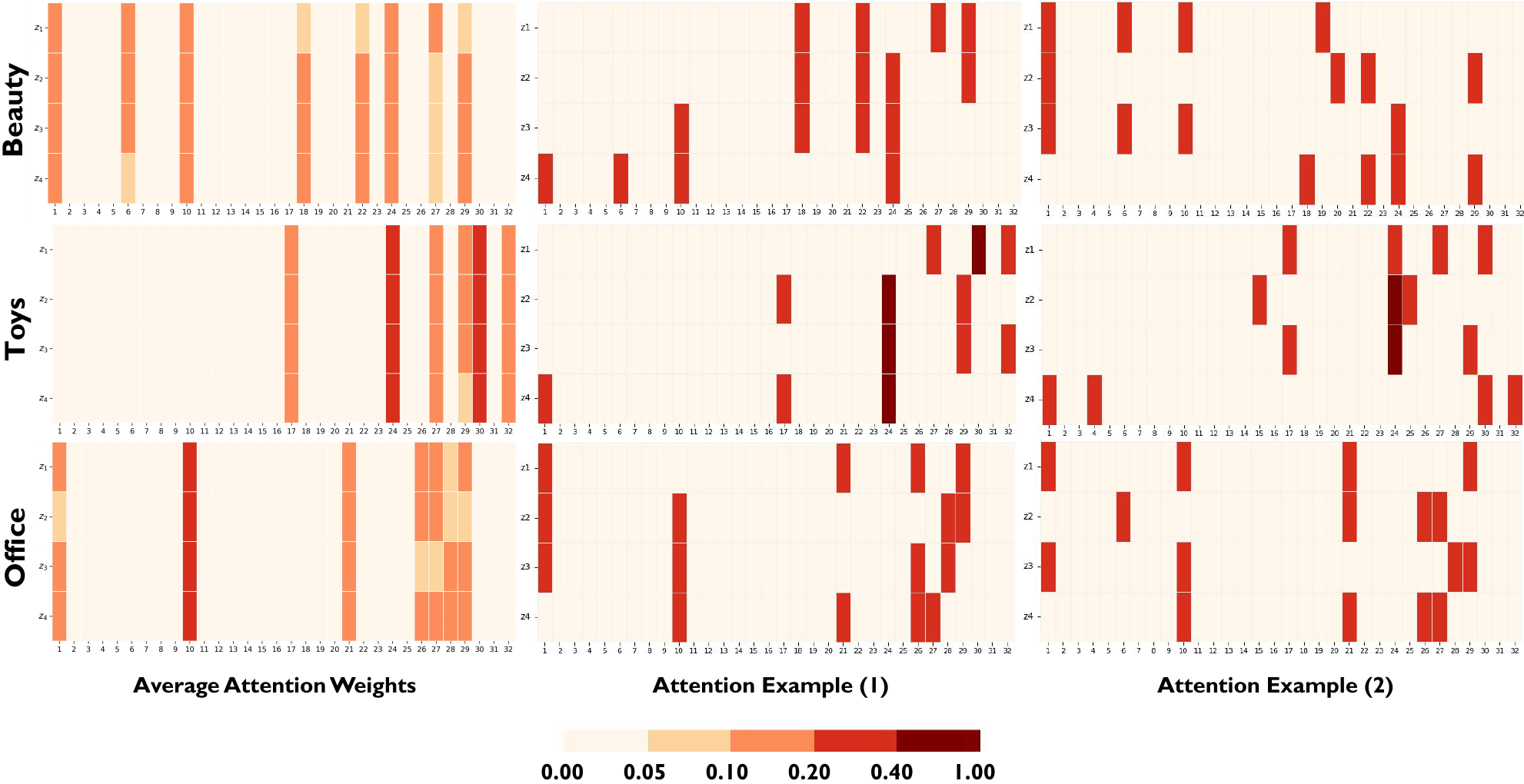}
    
    \caption{Visualization of ATM compression.}
    \label{Fig: atm_vis}
\end{figure*}

\subsection{Visualization of Adaptive Semantic Distillation} \label{app:atm_visualization}

Figure~\ref{Fig: atm_vis} visualizes the ATM attention weights, offering insights into how the model bridges high-fidelity inputs ($m=32$) and compact latent states ($k=4$).

First, the global average patterns (Col. 1) reveal a ``sparse-yet-focused'' activation mechanism. ATM automatically identifies domain-specific ``hotspot'' subspaces (e.g., indices 24 and 30 in Toys) that likely encode dominant attributes, while assigning near-zero weights to less informative regions. This confirms that ATM acts as a semantic distiller, extracting high-purity signals rather than performing rigid pooling and diluting distinctive features.

Second, the individual examples (Cols. 2 \& 3) demonstrate ATM's \textit{content-adaptive} nature. While adhering to the global structure, distinct items trigger different semantic subspaces. For instance, in Beauty, Example 1 focuses on index 27, whereas Example 2 shifts attention to index 19. This demonstrates that ATM performs dynamic, on-demand sampling: distinct items require access to different subspaces to be fully described. A static short code (e.g., $m=4$) would forcibly collapse these distinct features, leading to attribute entanglement.

\end{document}